\documentclass[aps,pra,reprint,amsmath, amssymb,
superscriptaddress,showpacs]{revtex4-2}
\usepackage[colorlinks = true, linkcolor = blue, urlcolor  = blue, citecolor = blue, anchorcolor = blue]{hyperref}
\usepackage{xspace}
\usepackage{multirow, amsmath, array, booktabs, color, bm}
\usepackage{longtable,mathrsfs}
\usepackage{ epsfig, latexsym, amssymb}
\usepackage[section]{placeins}
\usepackage[textsize=footnotesize]{todonotes}
\usepackage{dcolumn}
\usepackage{physics}

\usepackage{graphicx}
\usepackage[caption=false]{subfig}
\newcommand{\prlsection}[1]{\textit{#1.—}}

\begin{document}

\title{Super-Heisenberg-limited Sensing via Collective Subradiance in Waveguide Quantum Electrodynamics}

\author{Xin Wang}
\affiliation{School of Physics, Sun Yat-sen University, Guangzhou 510275, China}
\author{Zeyang Liao}
\email[E-mail:]{liaozy7@mail.sysu.edu.cn}
\affiliation{School of Physics, Sun Yat-sen University, Guangzhou 510275, China}

\begin{abstract}
    We explore the quantum-metrological potential of subwavelength-spaced emitter arrays coupled to a one-dimensional nanophotonic waveguide. In this system, strong dipole--dipole interactions profoundly modify the collective optical response, leading to the emergence of ultranarrow subradiant resonances. Through an eigenmode analysis of the effective non-Hermitian Hamiltonian, we derive the decay rate of the most subradiant state, which, within the present model and geometry, exhibits an $N^{-3}$ scaling with even--odd oscillatory behavior in the deep-subwavelength regime. This scaling is directly observable in the single-photon scattering spectrum, enabling the detection of minute changes in atomic separation with a figure of merit that scales as $ N^3 $. The quantum Fisher information (QFI) scales as $N^6$ and can be closely approached by measuring spectral shifts near the steepest slope of the most subradiant resonance. These enhancements remain robust under realistic positional disorder, confirming that dipole--dipole-engineered subradiance provides a viable resource for quantum metrology. Our work bridges collective waveguide-QED physics and high-precision sensing, opening a route toward scalable quantum sensors on integrated nanophotonic platforms.
\end{abstract}
\maketitle

\section{Introduction}
Quantum metrology aims to exploit quantum resources to measure physical parameters with a precision surpassing the classical shot-noise (standard quantum) limit (SQL), where the estimation error scales as $1/\sqrt{N}$ with the number $N$ of independent probes
~\cite{giovannetti2004, giovannetti2011,degen2017,pezze2018,liao2024}. 
In linear interferometry with entangled states such as NOON states, the Heisenberg limit (HL) with $\propto 1/N$ scaling represents the ultimate scaling allowed by {linear unitary evolution under local generator and noiseless conditions}
~\cite{dowling2008,paris2009,braunstein1994,xie2021,colombo2022a,qiu2022,zhang2023f,liu2021}. However, practical realizations based on entangled states face severe challenges from decoherence, photon loss, and the complexity of preparing large-scale multiphoton entangled states 
\cite{zhang2015a,munozdelasheras2024}. A promising new approach is critical quantum metrology, which exploits the enhanced susceptibility and nonclassical correlations found near quantum phase transitions to achieve quantum-enhanced precision 
\cite{zanardi2008,frerot2018,chu2021,ding2022,liu2021}, 
but critical slowing down poses a major challenge for the practical implementation of this scheme
 \cite{schneider1972,brookes2021}.
Nonlinear quantum metrology leverages higher-order probe–system couplings or many-body collective effects to achieve super-Heisenberg scaling $N^{-\alpha}$ with $\alpha>1$, 
as predicted for systems with $k$-body interactions where $\alpha = k$ in the ideal case
~\cite{luis2004,boixo2008,boixo2007,napolitano2011,Montenegro2025}. Nevertheless, achieving interactions beyond two-body remains challenging in realistic physical systems.

The development of integrated quantum technologies has spurred significant interest in waveguide quantum electrodynamics (waveguide-QED), a platform for studying the interaction between quantum emitters and a one-dimensional photonic continuum 
~\cite{domokos2002,shen2005a,liao2010,liao2016a,sheremet2023,roy2017,tian2025}. 
A pivotal frontier in this field is the exploration of many-body physics, where multiple emitters, collectively coupled to the waveguide, exhibit exotic correlated states such as superradiance and subradiance due to long-range interactions mediated by guided photons 
~\cite{zeyangliao2015,goban2015,perczel2017,cheng2017,song2017,chang2018a,fayard2021a,reitz2022,nie2023,lu2024}.
While these collective phenomena have been harnessed for applications like quantum information processing 
~\cite{paulisch2016,albrecht2019,bluvstein2022,masson2020,zhang2019a,shahmoon2017,rui2020,xing2024a,xing2024,lu2025,gonzalez-tudela2015,gonzalez-tudela2017} 
and photon manipulation 
~\cite{ke2019,wang2024b,wang2025,lu2025},
the potential of subradiant states—characterized by their extremely low decay rates—for quantum sensing remains largely untapped 
~\cite{liao2017,zhou2025}.

In this work, we bridge this gap by establishing a fundamental connection between subradiance and ultra-sensitive metrology. We derive an analytical expression for the decay rate of the most subradiant state in a system where atoms couple to both waveguide and unguided modes. Our results show that, within the present model and geometry, the decay rate follows an $N^{-3}$ scaling, consistent with previous predictions in the limiting cases of an ideal waveguide or free space~\cite{asenjo-garcia2017, zhang2020b, kornovan2019}.
  {Notably, we reveal that in the deep-subwavelength regime, the decay rate exhibits oscillations with respect to the number of atoms, a behavior attributable to finite-size effects}. Building on this insight, we propose a sensing scheme that leverages this scaling to achieve a quantum enhancement where the quantum Fisher information scales with $N^{6}$, far exceeding the Heisenberg limit {due to the long-range interaction and nonlinear encoding of the measuring parameters}.  More importantly, we find that this quantum limit can be nearly saturated by a straightforward classical measurement scheme---namely, detecting intensity variations on the steepest slope of the most subradiant resonance spectrum.
  
In contrast to approaches based on highly entangled multiphoton states, our scheme operates with a single incident photon, requires no active entanglement generation, and relies only on linear optics and straightforward reflection/transmission measurements.
Moreover, 
by exploiting the integrated nature of waveguide-QED platforms, our scheme constitutes a highly promising architecture for a new generation of compact, robust, and quantum-enhanced sensors.

The organization of this paper is as follows. The theoretical model adopted in this work, along with the decay rates of the most subradiant states for both ideal and non-ideal waveguide-QED configurations, is presented in Sec. II. The use of subradiant spectral shifts for separation sensing is elaborated in Sec. III. In Sec. IV, the quantum Fisher information is computed to quantify the ultimate sensitivity of the proposed scheme.  Possible implementation platforms are discussed in Sec. V. Finally, the paper concludes with a summary of our principal results.

\section{Theoretical model}
\begin{figure}[htbp]
    \centering
        \includegraphics[width=0.95\linewidth]{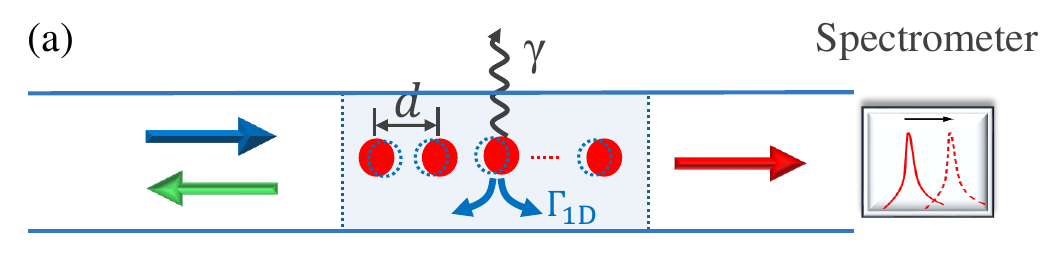} 
    \includegraphics[width=1\linewidth]{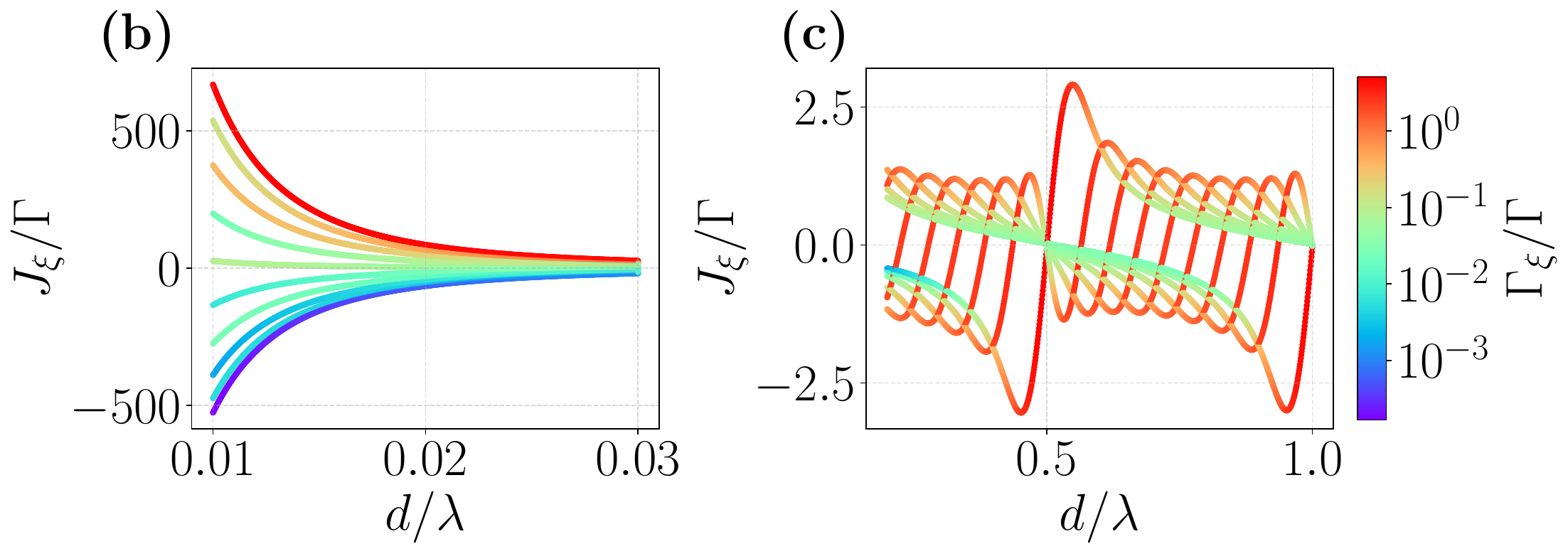} 
    \caption{ 
            (a) Schematic of a quantum sensor based on an array of $N$ two-level atoms coupled to a one-dimensional waveguide. 
            (b)(c) Eigenenergies as a function of atom separation $d$ when $\gamma = 0.1\Gamma$, shown for (b) smaller ($ d \in [0.01 \lambda, 0.03 \lambda] $) and (c) larger ($ d \in [0.12 \lambda, \lambda] $) separations with the color indicating the corresponding decay rates $\Gamma_\xi$ when $N=10$. 
}
    \label{fig:1}
\end{figure}

The schematic setup is shown in Fig.~\ref{fig:1}(a), where $ N $ identical two-level atoms with transition angular frequency $\omega_{0}$ couple to a 1D waveguide. All atoms are equally spaced with separation $d$ along the axis of a single-mode 1D waveguide, and the unguided modes are approximately treated as a free-space vacuum. After tracing out the photonic degrees of freedom, the effective Hamiltonian is given by
~\cite{liao2016,liao2017,asenjo-garcia2017}
\begin{equation} \label{eq:Heff1}
  \mathcal{H}_{\mathrm{eff}}=-i\sum_{j,l=1}^{N}\left[\frac{\Gamma}{2}\mathrm{e}^{ik_{\mathrm{1D}} z_{jl}}+V_{jl}\mathrm{e}^{ik_0 z_{jl}}\right]\hat{\sigma}_j^{+}\hat{\sigma}_l^{-},
\end{equation}
where $\Gamma$ is the decay rate due to the waveguide mode and $k_{\mathrm{1D}}$ is the guided-mode wave number evaluated at the atomic resonance. 
Here $V_{jl}\mathrm{e}^{ik_0 z_{jl}}$ denotes the effective dipole–dipole (DD) interaction mediated by unguided modes, with potential
$V_{jl}=(3\gamma/4)[-i/(k_0 z_{jl})+1/(k_0^2 z_{jl}^2)+i/(k_0^3 z_{jl}^3)]$ for dipoles $\mathbf{p}$ perpendicular to the waveguide axis, where $\gamma$ is the free-space decay rate, $V_{ii}=\gamma/2$, $k_0=\omega_{0}/c$ and $z_{jl}=|z_j-z_l|$ is the separation between the $j$-th and $l$-th emitters. The free-space dipole--dipole kernel used here assumes fixed emitter positions and linearly polarized transition dipoles perpendicular to the waveguide axis, corresponding to reciprocal guided coupling. Circular or elliptical dipoles may generate chiral, nonreciprocal guided interactions through spin--momentum locking; this extension is discussed in the Supplementary Material.  

Equation~\eqref{eq:Heff1} is a simplified effective Hamiltonian which is obtained after eliminating the photonic degrees of freedom under the Born--Markov and rotating-wave approximations. In this description, the guided channel is treated as the dominant structured long-range contribution of the waveguide environment, while the remaining nonguided modes are approximated as a weakly modified free-space reservoir~\cite{gonzalez-tudela2017}. More generally, the electromagnetic Green's tensor of the full structure may be decomposed as
$ \mathbf{G}(\mathbf{r},\mathbf{r}',\omega)=\mathbf{G}_{0}(\mathbf{r},\mathbf{r}',\omega)+\mathbf{G}_{s}(\mathbf{r},\mathbf{r}',\omega),$ 
where $\mathbf{G}_{0}$ is the free-space Green's tensor and $\mathbf{G}_{s}$ is the scattering contribution induced by the waveguide. In the effective model used here, the dominant long-range part of $\mathbf{G}_{s}$ is approximated by the single guided mode, whereas the residual radiation channels are represented by the free-space reservoir. This approximation is appropriate when the waveguide is close to the single-mode regime around the atomic resonance, when higher-order modes are cut off or evanescent, and when imperfections that scatter guided photons into radiation modes are weak.  In addition, the atom-atom separation discussed in this work is still well above the sub-nanometer regime where electron exchange, orbital overlap, and other quantum-chemistry effects are expected to become important. In this parameter range, the near-field dipole--dipole interaction is enhanced but remains far below the bare optical transition frequency, so the single-excitation Markovian description remains appropriate. The model should not be extrapolated to ultrashort separations where nonradiative microscopic chemistry effects or a breakdown of the effective reservoir description may occur~\cite{andreoli2023}.

It is noted from Eq.~\eqref{eq:Heff1} that the parameter of interest in this work---the atomic separation---is nonlinearly encoded in the global and long-range interaction kernel of the dipolar atom array. The effective Hamiltonian $\mathcal{H}_{\mathrm{eff}}$ is a complex symmetric matrix that can be block-diagonalized by exploiting the conservation of the total excitation number. In the single-excitation subspace, solving the eigenvalue equation $\mathcal{H}_{\text{eff}} |\phi_\xi\rangle = \lambda_\xi |\phi_\xi\rangle$ yields $N$ eigenstates where $\xi = 1, \dots, N$. The eigenvalues are $\lambda_\xi= J_\xi - i \Gamma_\xi/2$,  where $J_\xi = \mathrm{Re}(\lambda_\xi)$ represents the frequency shift and $\Gamma_\xi = -2\mathrm{Im}(\lambda_\xi)$ denotes the decay rate of the corresponding collective emitter mode
~\cite{a.asenjo-garcia2017,wang2025}. 
The system's collective excited states are described by eigenvectors of the form $|\phi_\xi\rangle=\sum_{j=1}^{N}c_{\xi}(j)|e_j\rangle$, where $|e_j\rangle$ denotes the state with only the $j$-th atom excited. The coefficients are normalized such that $\sum_{j=1}^{N}|c_{\xi}(j)|^2=1$. 
These eigenmodes can be categorized as either superradiant ($ \Gamma_{\xi} > \Gamma $) or subradiant ($ \Gamma_{\xi} < \Gamma $). We focus on the most subradiant states with the smallest $ \Gamma_{\xi} $, which provide a rich resource for sensitive quantum metrology. In the discussion that follows, we adopt the approximation $k_{\mathrm{1D}}\approx k_0$ for simplicity. This reduces the number of independent phase parameters, thereby enabling us to focus on the collective subradiance mechanism. Results for the case $k_{\mathrm{1D}}\neq k_0$ are analogous to those obtained under the approximation $k_{\mathrm{1D}}\approx k_0$ (for a detailed comparison, please refer to the Supplementary Material).

For an ideal waveguide in the absence of external dissipation ($\gamma=0$), the decay rates of the subradiant collective modes were derived by Zhang \textit{et al.}~\cite{zhang2019a}. 
{For completeness, we provide a detailed re-derivation in Sec. I of the supplementary material~\cite{supplementary}}:
\begin{equation}
\Gamma_{\xi}(N,d)\approx\frac{\Gamma}{2}\,\frac{\pi^2\xi^2}{N^3}\,
\frac{\sin^{2}(k_{0}d/2)}{\cos^{4}(k_{0}d/2)} ,
\label{eq:Gamma_zhang}
\end{equation}
where $\xi=1$ corresponds to the most subradiant mode (i.e., the smallest collective decay rate). When the guided-mode wave number differs from the free-space wave number, the guided-mode phase in Eq.~\eqref{eq:Gamma_zhang} should be read as $k_{\mathrm{1D}}d$ rather than $k_0d$.

For nonideal waveguides, when $d$ is sufficiently large (approximately ($ d \gtrsim \lambda/10 $)) and the unguided decay rate satisfies $\gamma \ll \Gamma$, Eq.~\eqref{eq:Gamma_zhang} remains valid. In this regime, the guided modes still dominate the collective dipole--dipole interaction, and contributions from unguided modes are negligible. As Fig.~\ref{fig:1}(c) illustrates, the most subradiant states lie near the band center. Including both guided and nonguided channels is nevertheless important for nonideal waveguides, because fabrication imperfections, finite mode confinement, and radiation loss generally make the two reservoirs coexist.

When the nonideal waveguides are in the deep-subwavelength regime ($d \ll \lambda$), atom-atom interactions are now dominated by unguided free-space modes. Figure~\ref{fig:1}(b) demonstrates that the eigenstates diverge rapidly away from the band center as $ d $ shrinks, and the most subradiant states occur at small $J_\xi$. Consequently, Eq.~\eqref{eq:Gamma_zhang} is no longer applicable and must be modified.

While the non-Hermitian nature of $\mathcal{H}_{\mathrm{eff}}$ generally leads to nonorthogonal right eigenvectors $|\phi_{\xi}\rangle$, the most-subradiant manifold forms a notable exception: owing to their strongly suppressed radiative decay, these modes are nearly orthogonal to an excellent approximation. 
{For an open chain (i.e., with two free ends), the single-excitation eigenmodes near the Bragg edge are conveniently approximated by Dirichlet sine modes \cite{a.asenjo-garcia2017}
\begin{equation}
  c_{\xi}(j)\approx \sqrt{\frac{2}{N+1}}  \sin\!\Big(\frac{\pi\xi\,j}{N+1}\Big)\,
  e^{ik_\xi z_j},
  \label{eq:phi-OB}
\end{equation}
}
where $k_\xi$ labels the corresponding quasimomentum.

The collective decay rate can be written as
$\Gamma_{\xi}=-2\,\mathrm{Im}\!\left(\langle\phi_\xi|H_{\mathrm{eff}}|\phi_{\xi}\rangle\right)$,
and decomposed into guided and unguided contributions,
$\Gamma_{\xi}=\Gamma_{\xi}^{(\mathrm{1D})}+\Gamma_{\xi}^{(\mathrm{fs})}$.
Explicitly,
\begin{equation}
  \begin{aligned}
    &\Gamma_{\xi}^{(\mathrm{1D})}=\Gamma\left|\sum_{j=1}^{N}c_{\xi}(j)e^{ik_{\mathrm{1D}} z_j}\right|^{2},\\
\qquad
&\Gamma_{\xi}^{(\mathrm{fs})}=\gamma \sum_{j,l=1}^{N}c_{\xi}^{*}(j)c_{\xi}(l)\, 
\mathcal{K}_{\mathrm{fs}}\!\big(k_0 |z_j-z_l|\big),
  \end{aligned}
\end{equation}
with
$\mathcal{K}_{\mathrm{fs}}(x)=\frac{3}{2}\!\left[\frac{\sin x}{x}+\frac{\cos x}{x^{2}}-\frac{\sin x}{x^{3}}\right]$.
The most-subradiant mode typically resides near the band edge ($k_\xi\simeq \pm\pi/d$), where destructive interference between emissions from different atoms suppresses radiation into both guided and unguided channels. 
Although a closed-form expression for $\Gamma_{\xi}$ is not available for all $ d $, in the deep-subwavelength limit ($d\ll\lambda$) one can obtain (see the derivation details from Sec. II in ~\cite{supplementary})
\begin{equation}
    \begin{aligned}
        \Gamma_\xi(N,d)\;&\approx \;\frac{\pi^2\xi^2}{(N+1)^3}
\big\{\frac{\Gamma}{4}[1+(-1)^{N+\xi}\text{cos}(\theta_{N+1})] \\
&+\frac{\gamma}{4}[1+(-1)^{N+\xi}\mathcal{K}_{\rm fs}(\theta_{N+1})]\big\},
    \end{aligned}
\label{eq:Gamma_final}
\end{equation}
where $\theta_{N+1}=(N+1)k_0 d$.

Equation~\eqref{eq:Gamma_final} shows that, even for a nonideal waveguide, the decay rate of the most-subradiant mode retains the overall $N^{-3}$ scaling, but splits into two interwoven branches for even and odd $N$. 
This parity effect originates from boundary-induced interference encoded by the $(-1)^{N+1}$ terms: incrementing $N\to N+1$ flips the sign of the interference contribution, thereby alternating the linewidth between the two branches.

\begin{figure}[htbp]
    \centering
     \includegraphics[width=1\linewidth]{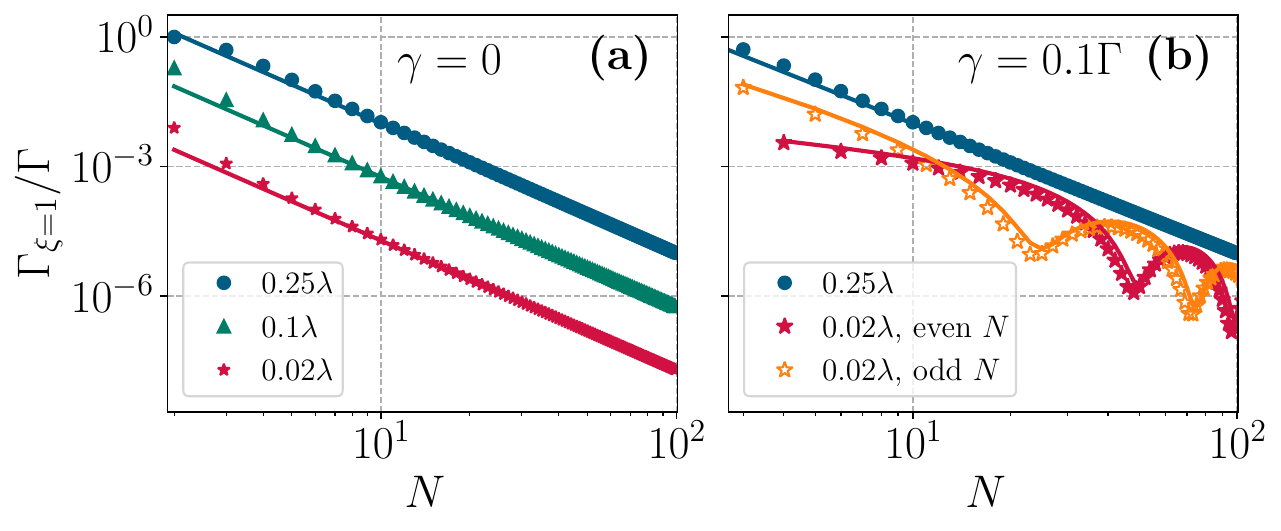} 
    \caption{
        Most-subradiant decay rate $\Gamma_{\xi=1}/\Gamma$ versus atom number $N$.
        (a) Ideal waveguide ($\gamma=0$) for three lattice spacings: $d=0.25\lambda$ (circles), $d=0.10\lambda$ (triangles), and $d=0.02\lambda$ (stars). Solid lines are the analytical prediction of Eq.~(\ref{eq:Gamma_zhang}).
        (b) Nonideal waveguide ($\gamma=0.1\Gamma$) for two lattice spacings: $d=0.25\lambda$ (circles) and $d=0.02\lambda$ (stars). For $d=0.02\lambda$, the linewidth splits into two branches: filled stars (even $N$) and open stars (odd $N$); solid curves given by Eq.~(\ref{eq:Gamma_final}) capture both branches.
        Symbols are numerical results.
            }
        \label{fig:2}
\end{figure}

To assess the validity of Eqs.~\eqref{eq:Gamma_zhang} and~\eqref{eq:Gamma_final}, we numerically diagonalize the effective non-Hermitian Hamiltonian $\mathcal{H}_{\mathrm{eff}}$ in Eq.~\eqref{eq:Heff1} for both ideal and nonideal waveguides. Figures~\ref{fig:2}(a) and~\ref{fig:2}(b) show the decay rate of the most-subradiant mode, $\Gamma_{\xi=1}$, as a function of atom number $N$ for several lattice spacings $d$, for $\gamma=0$ and $\gamma=0.1\Gamma$, respectively. Symbols represent the numerical eigenvalues, while solid curves are evaluated from the analytical expressions, Eq.~\eqref{eq:Gamma_zhang} for panel (a) and Eq.~\eqref{eq:Gamma_final} for panel (b).

For an ideal waveguide, $\Gamma_{\xi=1}$ follows the expected inverse-cubic scaling with atom number, $\Gamma_{\xi=1}\propto N^{-3}$, in excellent agreement with Eq.~\eqref{eq:Gamma_zhang}. Moreover, at fixed $N$, decreasing the lattice spacing $d$ further suppresses $\Gamma_{\xi=1}$, reflecting stronger destructive interference in the radiation from different emitters.

For a nonideal waveguide ($\gamma\neq 0$), interactions and decay via unguided modes contribute to the collective linewidth. In the moderately subwavelength regime (e.g., $d=0.25\lambda$), these contributions are small compared with the guided-mode-mediated coupling, so $\Gamma_{\xi=1}$ closely resembles the ideal-waveguide behavior and remains well described by Eq.~\eqref{eq:Gamma_zhang}. In contrast, in the deep-subwavelength regime ($d\ll\lambda$), $\Gamma_{\xi=1}$ develops a pronounced even--odd staggering with $N$, which is quantitatively captured by Eq.~\eqref{eq:Gamma_final}. Despite this parity-induced splitting, the envelope of the linewidth continues to decrease approximately as $N^{-3}$.

\section{Separation sensing via subradiant spectral shifts} 
Consider a single photon of angular frequency $\omega$ incident from one end of the waveguide. 
The transmittance is given by Refs.~\cite{liao2016,liao2017}:
\begin{equation}\label{eq:T}
T(\omega,d)=\left|1-\frac{\Gamma}{2}\sum_{j,l=1}^{N}M_{jl}^{-1}(\omega,d)\,e^{i\omega(z_{l}-z_{j})/c}\right|^2 ,
\end{equation}
where $M^{-1}$ is the inverse of the coupling matrix $M$ with matrix elements
$M_{jl}(\omega,d)=(\Gamma/2+V_{jl})e^{i\omega z_{jl}/c}-i(\omega-\omega_{0})\delta_{jl}$.
In the lossless limit ($\gamma=0$), one has the reflectance $R(\omega,d)=1-T(\omega,d)$; 
{for $\gamma\neq0$, generally $T(\omega,d)+R(\omega,d)<1$, and the missing fraction corresponds to loss into unguided modes.} 
The spectral features—particularly the frequency positions and linewidths (FWHMs) of peaks or dips—are well captured by the frequency shifts $ J_{\xi} $ and decay rates $ \Gamma_{\xi} $. This correspondence is especially pronounced for subradiant modes due to their narrow linewidths.

\begin{figure}[!htbp]
    \centering 
    \includegraphics[width=1\linewidth]{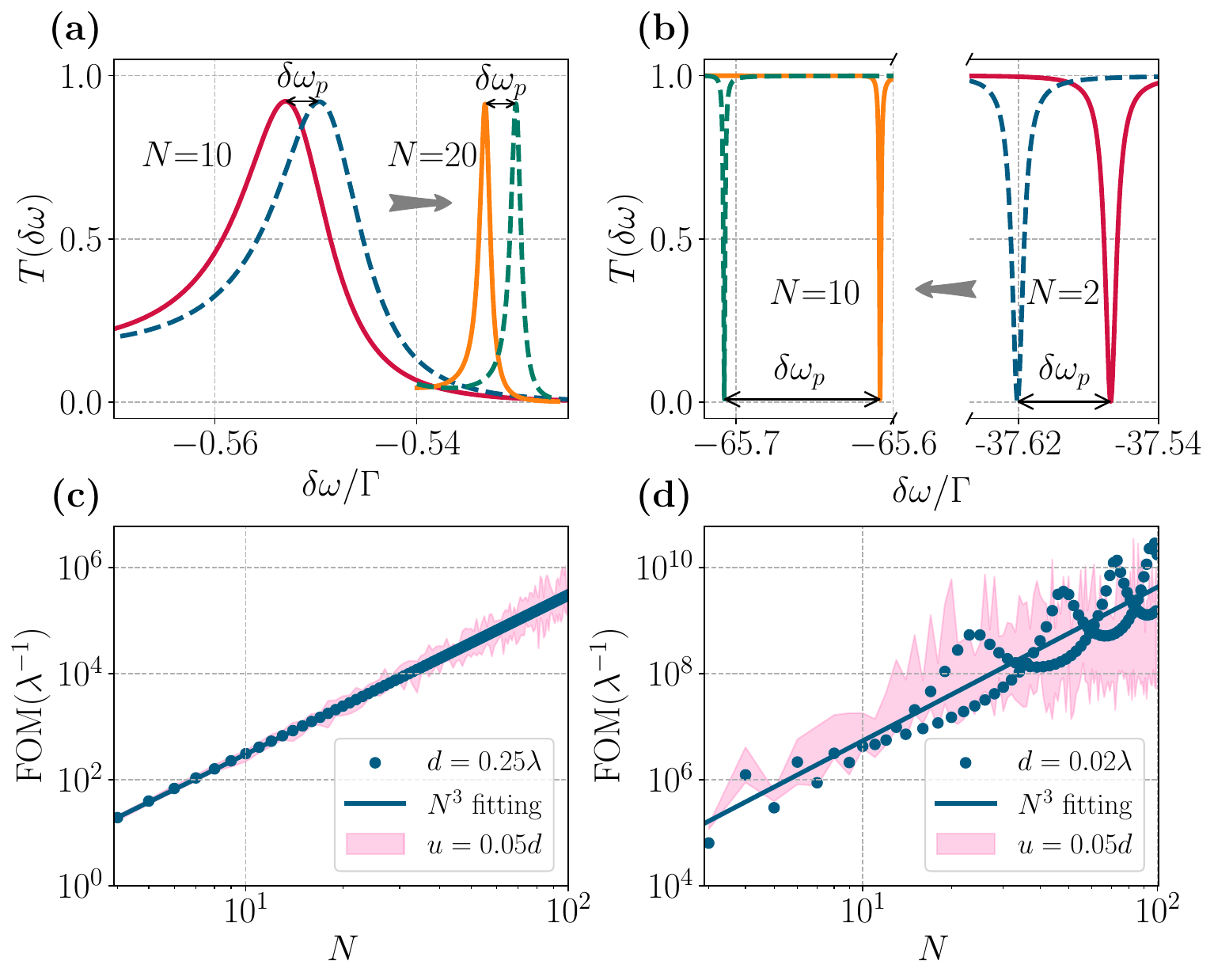} 
    \caption{
            (a) Evolution of transmission spectra near a subradiant feature as the atom number increases from $N=10$ to $N=20$. Solid and dashed curves correspond to inter-emitter spacings $d=0.25\lambda$ and $d'=0.249\lambda$, respectively.
            (b) Spectral shifts of the most subradiant dip for $d=0.02\lambda$ (solid) and $d'=0.01999\lambda$ (dashed) as $N$ increases from $2$ to $10$.
            (c)(d) Figure of merit (FOM) for the subradiant peak at (c) $d=0.25\lambda$ and the most subradiant dip at (d) $d=0.02\lambda$ as a function of $N$. Solid lines represent fits to an $N^3$ scaling. The pink bands indicate the FOM variation over 20 disorder realizations with a positional amplitude $u=0.05d$. Parameters: $\gamma=0.1\Gamma$.
    }
    \label{fig:3}
\end{figure}

Given that realistic waveguides are typically nonideal ($\gamma\neq 0$), we focus on this general case. 
We quantify separation sensing by tracking the frequency $\omega_p$ of the selected most subradiant transmission feature under a small perturbation $\delta d$. 
Specifically, we consider the spectral responses of the most subradiant modes for $d=0.25\lambda$ and $d=0.02\lambda$ in Figs.~\ref{fig:3}(a) and~\ref{fig:3}(b).

For moderately subwavelength spacing ($d=0.25\lambda$), the subradiant and superradiant eigenfrequencies remain relatively clustered (Fig.~\ref{fig:1}(c)). This allows interference between the most superradiant mode and the most subradiant modes to produce sharp reflection minima (equivalently, transmission peaks, Fig.~\ref{fig:3}(a)). 
For a perturbation $\delta d = 10^{-3}\lambda$, the peak shifts by $\delta\omega_{p}=0.0033\Gamma$ for $N=10$ and $\delta\omega_{p}=0.00306\Gamma$ for $N=20$. 
While the absolute shifts are comparable, the linewidth for $N=20$ is substantially narrower than that for $N=10$, making the frequency shift easier to resolve.

In contrast, in the deep-subwavelength regime ($d\ll\lambda$), the strong, nonguided dipole–dipole interaction significantly spreads the eigenvalues, effectively isolating each resonance (Fig.~\ref{fig:1}(b)). The interference between the subradiant modes is negligible. Consequently, each subradiant resonance gives rise to a pronounced reflection peak, manifesting as a transmission dip (Fig.~\ref{fig:3}(b)).
Here, as Fig.~\ref{fig:3}(b) indicates, a much smaller perturbation $\delta d = 10^{-5}\lambda$ from $d=0.02\lambda$ induces a dip shift of $0.18233\Gamma$ for $N=2$ and $0.37318\Gamma$ for $N=10$. Again, the linewidth for $N=10$ is much narrower than that for $N=2$. 
In both cases, the progressive linewidth suppression with increasing $N$, together with an appreciable spectral shift, highlights the potential of collective subradiant resonances for high-sensitivity separation sensing.

The figure of merit (FOM) is defined as the ratio of the spectral shift sensitivity to the resonance linewidth to quantify the sensing performance
\begin{equation}
\text{FOM}=\frac{|\partial \omega_{p}/\partial d|}{\sigma_\text{FWHM}},
\end{equation}
where $\omega_p$ is the frequency of the most subradiant feature (peak or dip) and $\sigma_\text{FWHM}$ is its full width at half maximum. This metric directly reflects the resonance's sharpness and susceptibility: a higher FOM enables the detection of smaller perturbations, by producing measurable spectral shifts relative to a narrow linewidth.

The FOM as a function of atomic number $N$ is shown in Figs.~\ref{fig:3}(c) and~\ref{fig:3}(d) for $d=0.25\lambda$ and $d=0.02\lambda$, respectively. For $d=0.25\lambda$, the FOM follows a clear $N^3$ scaling. A similar cubic scaling, albeit superimposed with pronounced oscillations, is observed for $d=0.02\lambda$. This oscillatory behavior is well-captured by the decay rate model in Eq.~\eqref{eq:Gamma_final}. Notably, the FOM for $d=0.02\lambda$ is approximately four orders of magnitude larger than that for $d=0.25\lambda$, highlighting the superior sensing performance achievable at deep-subwavelength separations. These results underscore the exceptional potential of collective subradiant resonances for quantum sensing applications.

In addition, we also investigate the robustness of the collective subradiant modes and their associated FOM against positional disorder, which typically arises from fabrication inaccuracies or spatial inhomogeneities. To model this imperfection, the position of the $j$-th atom is given by $r_j = jd + u \cdot \epsilon_j$, where $\epsilon_j$ is a uniformly distributed random variable in $[-1, 1]$, and $u$ denotes the disorder amplitude, representing a maximum positional deviation of $\pm u$.

Using a representative disorder strength of $u = 0.05d$, the resulting FOM is illustrated by the pink shaded bands in Figs.~\ref{fig:3}(c) and~\ref{fig:3}(d). These results indicate that the FOM remains notably robust under positional disorder. For the array with $d = 0.25\lambda$ [Fig.~\ref{fig:3}(c)], the disorder-induced variance in FOM is narrow, and the clean system's $N^3$ scaling is largely preserved across all atom numbers $N$. In the deep-subwavelength case with $d = 0.02\lambda$ [Fig.~\ref{fig:3}(d)], where unguided dipole-dipole interactions dominate, the FOM shows broader statistical fluctuations. Despite this, the overall $N^3$ scaling persists, and the absolute FOM values remain significantly high, underscoring the practical relevance of the proposed sensing scheme under realistic conditions.

\section{Fisher Information}
To quantify the ultimate precision in estimating the lattice spacing $d$, we calculate the Fisher information (FI) of the output scattering state of the system. For a single-photon scattering, the output photonic state in the waveguide is given by $|\psi_{\mathrm{out}}\rangle=[r(\omega,d)|R_\omega\rangle+t(\omega,d)|T_\omega\rangle]/\sqrt{p_{g}(\omega,d)}$, where $|R_\omega\rangle$ and $|T_\omega\rangle$ are the reflection and transmission photonic state with corresponding coefficients $r(\omega,d)$ and $t(\omega,d)$. The parameter $p_{g}(\omega,d)=|r(\omega,d)|^2+|t(\omega,d)|^2$ is the probability that the photon remains in the waveguide. The quantum Fisher information (QFI) is defined by 
~\cite{braunstein1994,Polino2020}
\begin{equation}
F_Q(|\psi_{\mathrm{out}}\rangle) = 4 \left( \langle \partial_d \psi_{\mathrm{out}} | \partial_d \psi_{\mathrm{out}} \rangle - |\langle \psi_{\mathrm{out}} | \partial_d \psi_{\mathrm{out}} \rangle|^2 \right)
.
\label{eq:QFI-guided}
\end{equation}

In addition to the QFI, we also calculate the classical FI via measuring the shift of the transmission spectrum at its steepest slope~\cite{degen2017,pezze2018}
\begin{equation}
F_{\mathrm{MT}}(d)
= \max_{\omega} \left\{ \frac{1}{T(\omega,d)} \left[ \frac{\partial T(\omega,d)}{\partial d} \right]^2 \right\}.
\label{eq:MT-max}
\end{equation}

\begin{figure}[!htbp]
    \centering
\includegraphics[width=0.75\linewidth]{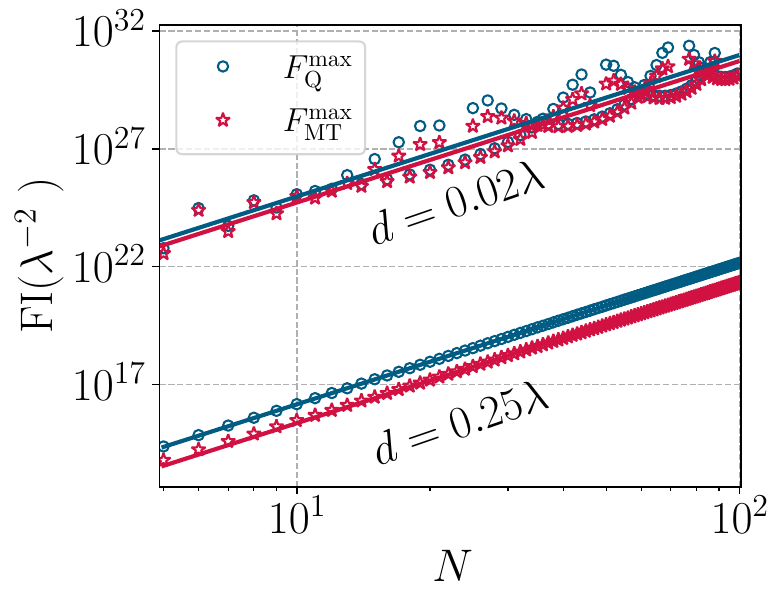} 
    \caption{
    Fisher information as a function of $N$ for $d = 0.25\lambda$ and $d = 0.02\lambda$. Blue circles: QFI; Red stars: FI.  Solid lines indicate $N^6$ scaling. Parameter: $\gamma = 0.1\Gamma$. 
    }
    \label{fig:4}
\end{figure}

In Fig.~\ref{fig:4}, we compare the QFI and FI as functions of $N$ for the cases $d = 0.25\lambda$ and $d = 0.02\lambda$. In both scenarios, the QFI and FI scale as $N^6$, which far exceeds the traditional Heisenberg limit. {Here, the observed super-linear scaling of the QFI arises from the nonlinear encoding of the measuring parameter $d$ in the long-range atom-atom interaction which breaks the linear and local assumptions in deriving the standard Heisenberg limit \cite{Montenegro2025}.}  From Fig.~\ref{fig:4}, we also see that the Fisher information for $d = 0.02\lambda$ is approximately six orders of magnitude larger than those for $d = 0.25\lambda$. This indicates that a sensor operating in the deep-subwavelength regime can achieve a significantly higher sensitivity.
Notably, we observe that $F_{\mathrm{MT}}$ closely approaches $F_{\mathrm{Q}}$ in both cases, implying that measuring the shift of the transmission spectrum at the steepest slope of the most subradiant resonance nearly saturates the quantum Fisher information limit. Thus, the ultimate QFI scaling is not only a formal bound but can be approximately approached by a direct spectral measurement based on transmitted intensity.

For a quantum sensor, the minimum resolvable change in distance—a key figure of merit—is bounded by the Cram\'{e}r–Rao inequality:
\begin{equation}
\delta d \geq \frac{1}{\sqrt{M \cdot F(d)}},
\end{equation}
where $ M $ is the number of independent measurements (e.g., detected photons) and $ F(d) $ is the Fisher information for each detection. Taking $F(d)=F_Q$ gives the ultimate quantum bound, while taking $F(d)=F_{\mathrm{MT}}$ gives the precision attainable by the transmission-slope measurement considered above. The ultimate quantum bound can be attained by performing a measurement of the entire scattering quantum state, which encompasses both the reflection and transmission amplitudes and the associated phases. Assuming $ M = 100 $ and only shot noise is present, a resolution of $ \delta d < 10^{-12} \lambda $ is achievable with $ N = 100 $ atoms at a separation of $ d = 0.25\lambda $.  In contrast, a comparable precision at $ d = 0.02\lambda $ requires only $ N = 10 $ atoms, underscoring the superior sensitivity in the deep-subwavelength regime. Therefore, provided that technical noise sources are sufficiently suppressed, a resolution well below the picometer can in principle be resolved. Such a resolution scale is relevant for emerging applications such as nanomechanical sensing, strain and temperature detection, and refractive index sensing.


\section{Possible Physical Implementation}
One possible candidate for implementation of our proposal is a one-dimensional array of ultracold atoms, such as $^{87}$Rb or $^{133}$Cs, trapped in optical tweezers generated by a spatial light modulator or an optical metasurface 
~\cite{corzo2019,sunami2025}. 
The atoms can be coupled to the evanescent field of a nearby nanofiber or an integrated photonic crystal waveguide \cite{zhou2025selective,takahata2026fiber}. To mitigate position fluctuations and achieve the requisite coherence, atoms must be cooled to their motional ground state within the traps. Advanced cooling techniques, alongside the inherent stability of optical lattices used for ancillary positioning, can suppress positional disorder to a level where the collective subradiant physics becomes observable. Nevertheless, given the limitations of current technology, this setup is at present capable of demonstrating only the case of moderate subwavelength atom-atom separation, rather than the deep subwavelength regime. With continued development of trapping techniques \cite{pache2025magic,holzinger2022cooperative,Seubert2025,Srakaew2023}, realization of the deep subwavelength regime may become feasible in the future.

For a fully integrated and scalable implementation, solid-state quantum emitters such as quantum dots and the negatively charged silicon-vacancy ($\text{SiV}^{-}$) center in diamond may be better suited~\cite{sipahigil2016,ngan2023,tzeng2024,day2022coherent,agarwal2023quantum}. These emitters feature exceptional spectral stability and narrow inhomogeneous broadening at cryogenic temperatures—prerequisites for maintaining collective coherence—and can be nanofabricated into deterministic arrays evanescently coupled to nanophotonic waveguides. With spectral homogeneity achievable via strain and electric field engineering, such platforms can closely approximate the optimal measurement strategy outlined in this work. More importantly, nanofabrication can in principle realize emitter arrays with geometric separations far below the optical wavelength, making the deep subwavelength regime potentially accessible with current technology, provided that spectral inhomogeneity and nonradiative losses are carefully controlled~\cite{lan2012ordering,chen2023ultrafast,tidjani2025threedimensional,katsumi2025hybrid}. Superconducting circuits offer an alternative route, where deep-subwavelength arrays are easier to realize due to the much larger resonant wavelength. Moreover, capacitive couplings with an effective near-field-like distance dependence can emulate part of the free-space interaction used in the present model~\cite{zanner2022darkstate}.

\section{Summary}
We have demonstrated that, within the present model of an open one-dimensional emitter array coupled to both guided and nonguided modes, the most subradiant decay rate
retains an overall $N^{-3}$ scaling, while developing an analytically predictable even--odd oscillatory pattern in the deep-subwavelength regime. More importantly, by harnessing collective dipole--dipole interactions, such arrays achieve exceptional sensitivity to environmental perturbations, enabling the detection of minute spacing changes through spectral shifts in transmission or reflection. The derived FOM increases as $N^3$, while the quantum Fisher information scales as $N^6$, surpassing the Heisenberg limit in its dependence on the atom number due to the nonlinear encoding of the measuring parameter in the long-range collective interaction. Our analysis further establishes the robustness of these scaling laws against positional disorder, underscoring the practical relevance of the proposed architecture.

Within the idealized shot-noise-limited model, the QFI analysis shows that a resolution of $\delta d < 10^{-12}\lambda$ can in principle be attained for the parameters under consideration. Nevertheless, in practical implementations, technical noise, frequency instability, motional fluctuations, and model imperfections may compromise the resolution—an aspect that requires further study. This work opens a viable pathway for integrating collective subradiant states into practical quantum metrology platforms, paving the way for the development of highly compact, noise-tolerant, and ultra-sensitive quantum sensors.

\prlsection{\label{sec:acknowledgments}Acknowledgments}
This work was supported by the National Key R\&D Program of China (Grant No. 2021YFA1400800), the Key Program of National Natural Science Foundation of China (Grant No. 12334017), Guangdong Provincial Quantum Science Strategic Initiative (Grant No. GDZX2505001 and GDZX2406001), and Guangdong Basic and Applied Basic Research Foundation (Grant No. 2026A1515011705).


\begin{thebibliography}{86}%
\makeatletter
\providecommand \@ifxundefined [1]{%
 \@ifx{#1\undefined}
}%
\providecommand \@ifnum [1]{%
 \ifnum #1\expandafter \@firstoftwo
 \else \expandafter \@secondoftwo
 \fi
}%
\providecommand \@ifx [1]{%
 \ifx #1\expandafter \@firstoftwo
 \else \expandafter \@secondoftwo
 \fi
}%
\providecommand \natexlab [1]{#1}%
\providecommand \enquote  [1]{``#1''}%
\providecommand \bibnamefont  [1]{#1}%
\providecommand \bibfnamefont [1]{#1}%
\providecommand \citenamefont [1]{#1}%
\providecommand \href@noop [0]{\@secondoftwo}%
\providecommand \href [0]{\begingroup \@sanitize@url \@href}%
\providecommand \@href[1]{\@@startlink{#1}\@@href}%
\providecommand \@@href[1]{\endgroup#1\@@endlink}%
\providecommand \@sanitize@url [0]{\catcode `\\12\catcode `\$12\catcode
  `\&12\catcode `\#12\catcode `\^12\catcode `\_12\catcode `\%12\relax}%
\providecommand \@@startlink[1]{}%
\providecommand \@@endlink[0]{}%
\providecommand \url  [0]{\begingroup\@sanitize@url \@url }%
\providecommand \@url [1]{\endgroup\@href {#1}{\urlprefix }}%
\providecommand \urlprefix  [0]{URL }%
\providecommand \Eprint [0]{\href }%
\providecommand \doibase [0]{https://doi.org/}%
\providecommand \selectlanguage [0]{\@gobble}%
\providecommand \bibinfo  [0]{\@secondoftwo}%
\providecommand \bibfield  [0]{\@secondoftwo}%
\providecommand \translation [1]{[#1]}%
\providecommand \BibitemOpen [0]{}%
\providecommand \bibitemStop [0]{}%
\providecommand \bibitemNoStop [0]{.\EOS\space}%
\providecommand \EOS [0]{\spacefactor3000\relax}%
\providecommand \BibitemShut  [1]{\csname bibitem#1\endcsname}%
\let\auto@bib@innerbib\@empty
\bibitem [{\citenamefont {Giovannetti}\ \emph {et~al.}(2004)\citenamefont
  {Giovannetti}, \citenamefont {Lloyd},\ and\ \citenamefont
  {Maccone}}]{giovannetti2004}%
  \BibitemOpen
  \bibfield  {author} {\bibinfo {author} {\bibfnamefont {V.}~\bibnamefont
  {Giovannetti}}, \bibinfo {author} {\bibfnamefont {S.}~\bibnamefont {Lloyd}},\
  and\ \bibinfo {author} {\bibfnamefont {L.}~\bibnamefont {Maccone}},\
  }\bibfield  {title} {\bibinfo {title} {Quantum-enhanced measurements: Beating
  the standard quantum limit},\ }\href
  {https://doi.org/10.1126/science.1104149} {\bibfield  {journal} {\bibinfo
  {journal} {Science}\ }\textbf {\bibinfo {volume} {306}},\ \bibinfo {pages}
  {1330} (\bibinfo {year} {2004})}\BibitemShut {NoStop}%
\bibitem [{\citenamefont {Giovannetti}\ \emph {et~al.}(2011)\citenamefont
  {Giovannetti}, \citenamefont {Lloyd},\ and\ \citenamefont
  {Maccone}}]{giovannetti2011}%
  \BibitemOpen
  \bibfield  {author} {\bibinfo {author} {\bibfnamefont {V.}~\bibnamefont
  {Giovannetti}}, \bibinfo {author} {\bibfnamefont {S.}~\bibnamefont {Lloyd}},\
  and\ \bibinfo {author} {\bibfnamefont {L.}~\bibnamefont {Maccone}},\
  }\bibfield  {title} {\bibinfo {title} {Advances in quantum metrology},\
  }\href {https://doi.org/10.1038/nphoton.2011.35} {\bibfield  {journal}
  {\bibinfo  {journal} {Nat. Photon.}\ }\textbf {\bibinfo {volume} {5}},\
  \bibinfo {pages} {222} (\bibinfo {year} {2011})}\BibitemShut {NoStop}%
\bibitem [{\citenamefont {Degen}\ \emph {et~al.}(2017)\citenamefont {Degen},
  \citenamefont {Reinhard},\ and\ \citenamefont {Cappellaro}}]{degen2017}%
  \BibitemOpen
  \bibfield  {author} {\bibinfo {author} {\bibfnamefont {C.~L.}\ \bibnamefont
  {Degen}}, \bibinfo {author} {\bibfnamefont {F.}~\bibnamefont {Reinhard}},\
  and\ \bibinfo {author} {\bibfnamefont {P.}~\bibnamefont {Cappellaro}},\
  }\bibfield  {title} {\bibinfo {title} {Quantum sensing},\ }\href
  {https://doi.org/10.1103/RevModPhys.89.035002} {\bibfield  {journal}
  {\bibinfo  {journal} {Rev. Mod. Phys.}\ }\textbf {\bibinfo {volume} {89}},\
  \bibinfo {pages} {035002} (\bibinfo {year} {2017})}\BibitemShut {NoStop}%
\bibitem [{\citenamefont {Pezz{\`e}}\ \emph {et~al.}(2018)\citenamefont
  {Pezz{\`e}}, \citenamefont {Smerzi}, \citenamefont {Oberthaler},
  \citenamefont {Schmied},\ and\ \citenamefont {Treutlein}}]{pezze2018}%
  \BibitemOpen
  \bibfield  {author} {\bibinfo {author} {\bibfnamefont {L.}~\bibnamefont
  {Pezz{\`e}}}, \bibinfo {author} {\bibfnamefont {A.}~\bibnamefont {Smerzi}},
  \bibinfo {author} {\bibfnamefont {M.~K.}\ \bibnamefont {Oberthaler}},
  \bibinfo {author} {\bibfnamefont {R.}~\bibnamefont {Schmied}},\ and\ \bibinfo
  {author} {\bibfnamefont {P.}~\bibnamefont {Treutlein}},\ }\bibfield  {title}
  {\bibinfo {title} {Quantum metrology with nonclassical states of atomic
  ensembles},\ }\href {https://doi.org/10.1103/RevModPhys.90.035005} {\bibfield
   {journal} {\bibinfo  {journal} {Rev. Mod. Phys.}\ }\textbf {\bibinfo
  {volume} {90}},\ \bibinfo {pages} {035005} (\bibinfo {year}
  {2018})}\BibitemShut {NoStop}%
\bibitem [{\citenamefont {Liao}\ \emph {et~al.}(2024)\citenamefont {Liao},
  \citenamefont {Lu}, \citenamefont {Li},\ and\ \citenamefont
  {Wang}}]{liao2024}%
  \BibitemOpen
  \bibfield  {author} {\bibinfo {author} {\bibfnamefont {Z.}~\bibnamefont
  {Liao}}, \bibinfo {author} {\bibfnamefont {Y.-W.}\ \bibnamefont {Lu}},
  \bibinfo {author} {\bibfnamefont {W.}~\bibnamefont {Li}},\ and\ \bibinfo
  {author} {\bibfnamefont {X.-H.}\ \bibnamefont {Wang}},\ }\bibfield  {title}
  {\bibinfo {title} {Optical scattering imaging with sub-nanometer precision
  based on position-ultra-sensitive giant lamb shift},\ }\href
  {https://doi.org/10.1007/s11433-023-2369-6} {\bibfield  {journal} {\bibinfo
  {journal} {Sci. China Phys. Mech. Astron.}\ }\textbf {\bibinfo {volume}
  {67}},\ \bibinfo {pages} {264212} (\bibinfo {year} {2024})}\BibitemShut
  {NoStop}%
\bibitem [{\citenamefont {Dowling}(2008)}]{dowling2008}%
  \BibitemOpen
  \bibfield  {author} {\bibinfo {author} {\bibfnamefont {J.~P.}\ \bibnamefont
  {Dowling}},\ }\bibfield  {title} {\bibinfo {title} {Quantum optical
  metrology~--~the lowdown on high-{N00N} states},\ }\href
  {https://doi.org/10.1080/00107510802091298} {\bibfield  {journal} {\bibinfo
  {journal} {Contemp. Phys.}\ }\textbf {\bibinfo {volume} {49}},\ \bibinfo
  {pages} {125} (\bibinfo {year} {2008})}\BibitemShut {NoStop}%
\bibitem [{\citenamefont {Paris}(2009)}]{paris2009}%
  \BibitemOpen
  \bibfield  {author} {\bibinfo {author} {\bibfnamefont {M.~G.~A.}\
  \bibnamefont {Paris}},\ }\bibfield  {title} {\bibinfo {title} {Quantum
  estimation for quantum technology},\ }\href
  {https://doi.org/10.1142/S0219749909004839} {\bibfield  {journal} {\bibinfo
  {journal} {Int. J. Quantum Inform.}\ }\textbf {\bibinfo {volume} {07}},\
  \bibinfo {pages} {125} (\bibinfo {year} {2009})}\BibitemShut {NoStop}%
\bibitem [{\citenamefont {Braunstein}\ and\ \citenamefont
  {Caves}(1994)}]{braunstein1994}%
  \BibitemOpen
  \bibfield  {author} {\bibinfo {author} {\bibfnamefont {S.~L.}\ \bibnamefont
  {Braunstein}}\ and\ \bibinfo {author} {\bibfnamefont {C.~M.}\ \bibnamefont
  {Caves}},\ }\bibfield  {title} {\bibinfo {title} {Statistical distance and
  the geometry of quantum states},\ }\href
  {https://doi.org/10.1103/PhysRevLett.72.3439} {\bibfield  {journal} {\bibinfo
   {journal} {Phys. Rev. Lett.}\ }\textbf {\bibinfo {volume} {72}},\ \bibinfo
  {pages} {3439} (\bibinfo {year} {1994})}\BibitemShut {NoStop}%
\bibitem [{\citenamefont {Xie}\ \emph {et~al.}(2021)\citenamefont {Xie},
  \citenamefont {Zhao}, \citenamefont {Kong}, \citenamefont {Ma}, \citenamefont
  {Wang}, \citenamefont {Ye}, \citenamefont {Yu}, \citenamefont {Yang},
  \citenamefont {Xu}, \citenamefont {Wang}, \citenamefont {Wang}, \citenamefont
  {Shi},\ and\ \citenamefont {Du}}]{xie2021}%
  \BibitemOpen
  \bibfield  {author} {\bibinfo {author} {\bibfnamefont {T.}~\bibnamefont
  {Xie}}, \bibinfo {author} {\bibfnamefont {Z.}~\bibnamefont {Zhao}}, \bibinfo
  {author} {\bibfnamefont {X.}~\bibnamefont {Kong}}, \bibinfo {author}
  {\bibfnamefont {W.}~\bibnamefont {Ma}}, \bibinfo {author} {\bibfnamefont
  {M.}~\bibnamefont {Wang}}, \bibinfo {author} {\bibfnamefont {X.}~\bibnamefont
  {Ye}}, \bibinfo {author} {\bibfnamefont {P.}~\bibnamefont {Yu}}, \bibinfo
  {author} {\bibfnamefont {Z.}~\bibnamefont {Yang}}, \bibinfo {author}
  {\bibfnamefont {S.}~\bibnamefont {Xu}}, \bibinfo {author} {\bibfnamefont
  {P.}~\bibnamefont {Wang}}, \bibinfo {author} {\bibfnamefont {Y.}~\bibnamefont
  {Wang}}, \bibinfo {author} {\bibfnamefont {F.}~\bibnamefont {Shi}},\ and\
  \bibinfo {author} {\bibfnamefont {J.}~\bibnamefont {Du}},\ }\bibfield
  {title} {\bibinfo {title} {Beating the standard quantum limit under ambient
  conditions with solid-state spins},\ }\href
  {https://doi.org/10.1126/sciadv.abg9204} {\bibfield  {journal} {\bibinfo
  {journal} {Sci. Adv.}\ }\textbf {\bibinfo {volume} {7}},\ \bibinfo {pages}
  {eabg9204} (\bibinfo {year} {2021})}\BibitemShut {NoStop}%
\bibitem [{\citenamefont {Colombo}\ \emph {et~al.}(2022)\citenamefont
  {Colombo}, \citenamefont {{Pedrozo-Pe{\~n}afiel}}, \citenamefont
  {Adiyatullin}, \citenamefont {Li}, \citenamefont {Mendez}, \citenamefont
  {Shu},\ and\ \citenamefont {Vuleti{\'c}}}]{colombo2022a}%
  \BibitemOpen
  \bibfield  {author} {\bibinfo {author} {\bibfnamefont {S.}~\bibnamefont
  {Colombo}}, \bibinfo {author} {\bibfnamefont {E.}~\bibnamefont
  {{Pedrozo-Pe{\~n}afiel}}}, \bibinfo {author} {\bibfnamefont {A.~F.}\
  \bibnamefont {Adiyatullin}}, \bibinfo {author} {\bibfnamefont
  {Z.}~\bibnamefont {Li}}, \bibinfo {author} {\bibfnamefont {E.}~\bibnamefont
  {Mendez}}, \bibinfo {author} {\bibfnamefont {C.}~\bibnamefont {Shu}},\ and\
  \bibinfo {author} {\bibfnamefont {V.}~\bibnamefont {Vuleti{\'c}}},\
  }\bibfield  {title} {\bibinfo {title} {Time-reversal-based quantum metrology
  with many-body entangled states},\ }\href
  {https://doi.org/10.1038/s41567-022-01653-5} {\bibfield  {journal} {\bibinfo
  {journal} {Nat. Phys.}\ }\textbf {\bibinfo {volume} {18}},\ \bibinfo {pages}
  {925} (\bibinfo {year} {2022})}\BibitemShut {NoStop}%
\bibitem [{\citenamefont {Qiu}\ \emph {et~al.}(2022)\citenamefont {Qiu},
  \citenamefont {Zhuang}, \citenamefont {Huang},\ and\ \citenamefont
  {Lee}}]{qiu2022}%
  \BibitemOpen
  \bibfield  {author} {\bibinfo {author} {\bibfnamefont {Y.}~\bibnamefont
  {Qiu}}, \bibinfo {author} {\bibfnamefont {M.}~\bibnamefont {Zhuang}},
  \bibinfo {author} {\bibfnamefont {J.}~\bibnamefont {Huang}},\ and\ \bibinfo
  {author} {\bibfnamefont {C.}~\bibnamefont {Lee}},\ }\bibfield  {title}
  {\bibinfo {title} {Efficient and robust entanglement generation with deep
  reinforcement learning for quantum metrology},\ }\href
  {https://doi.org/10.1088/1367-2630/ac8285} {\bibfield  {journal} {\bibinfo
  {journal} {New J. Phys.}\ }\textbf {\bibinfo {volume} {24}},\ \bibinfo
  {pages} {083011} (\bibinfo {year} {2022})}\BibitemShut {NoStop}%
\bibitem [{\citenamefont {Zhang}\ \emph {et~al.}(2023)\citenamefont {Zhang},
  \citenamefont {Ye}, \citenamefont {Chang}, \citenamefont {Xia}, \citenamefont
  {Hu},\ and\ \citenamefont {Liao}}]{zhang2023f}%
  \BibitemOpen
  \bibfield  {author} {\bibinfo {author} {\bibfnamefont {H.}~\bibnamefont
  {Zhang}}, \bibinfo {author} {\bibfnamefont {W.}~\bibnamefont {Ye}}, \bibinfo
  {author} {\bibfnamefont {S.}~\bibnamefont {Chang}}, \bibinfo {author}
  {\bibfnamefont {Y.}~\bibnamefont {Xia}}, \bibinfo {author} {\bibfnamefont
  {L.}~\bibnamefont {Hu}},\ and\ \bibinfo {author} {\bibfnamefont
  {Z.}~\bibnamefont {Liao}},\ }\bibfield  {title} {\bibinfo {title} {Quantum
  multiparameter estimation with multi-mode photon catalysis entangled squeezed
  state},\ }\href {https://doi.org/10.1007/s11467-023-1274-6} {\bibfield
  {journal} {\bibinfo  {journal} {Front. Phys.}\ }\textbf {\bibinfo {volume}
  {18}},\ \bibinfo {pages} {42304} (\bibinfo {year} {2023})}\BibitemShut
  {NoStop}%
\bibitem [{\citenamefont {Liu}\ \emph {et~al.}(2021)\citenamefont {Liu},
  \citenamefont {Chen}, \citenamefont {Jiang}, \citenamefont {Yang},
  \citenamefont {Wu}, \citenamefont {Li}, \citenamefont {Yuan}, \citenamefont
  {Peng},\ and\ \citenamefont {Du}}]{liu2021}%
  \BibitemOpen
  \bibfield  {author} {\bibinfo {author} {\bibfnamefont {R.}~\bibnamefont
  {Liu}}, \bibinfo {author} {\bibfnamefont {Y.}~\bibnamefont {Chen}}, \bibinfo
  {author} {\bibfnamefont {M.}~\bibnamefont {Jiang}}, \bibinfo {author}
  {\bibfnamefont {X.}~\bibnamefont {Yang}}, \bibinfo {author} {\bibfnamefont
  {Z.}~\bibnamefont {Wu}}, \bibinfo {author} {\bibfnamefont {Y.}~\bibnamefont
  {Li}}, \bibinfo {author} {\bibfnamefont {H.}~\bibnamefont {Yuan}}, \bibinfo
  {author} {\bibfnamefont {X.}~\bibnamefont {Peng}},\ and\ \bibinfo {author}
  {\bibfnamefont {J.}~\bibnamefont {Du}},\ }\bibfield  {title} {\bibinfo
  {title} {Experimental critical quantum metrology with the heisenberg
  scaling},\ }\href {https://doi.org/10.1038/s41534-021-00507-x} {\bibfield
  {journal} {\bibinfo  {journal} {npj Quantum Inf}\ }\textbf {\bibinfo {volume}
  {7}},\ \bibinfo {pages} {170} (\bibinfo {year} {2021})}\BibitemShut {NoStop}%
\bibitem [{\citenamefont {Zhang}\ \emph {et~al.}(2015)\citenamefont {Zhang},
  \citenamefont {Mouradian}, \citenamefont {Wong},\ and\ \citenamefont
  {Shapiro}}]{zhang2015a}%
  \BibitemOpen
  \bibfield  {author} {\bibinfo {author} {\bibfnamefont {Z.}~\bibnamefont
  {Zhang}}, \bibinfo {author} {\bibfnamefont {S.}~\bibnamefont {Mouradian}},
  \bibinfo {author} {\bibfnamefont {F.~N.~C.}\ \bibnamefont {Wong}},\ and\
  \bibinfo {author} {\bibfnamefont {J.~H.}\ \bibnamefont {Shapiro}},\
  }\bibfield  {title} {\bibinfo {title} {Entanglement-enhanced sensing in a
  lossy and noisy environment},\ }\href
  {https://doi.org/10.1103/PhysRevLett.114.110506} {\bibfield  {journal}
  {\bibinfo  {journal} {Phys. Rev. Lett.}\ }\textbf {\bibinfo {volume} {114}},\
  \bibinfo {pages} {110506} (\bibinfo {year} {2015})}\BibitemShut {NoStop}%
\bibitem [{\citenamefont {{Mu{\~n}oz de las Heras}}\ \emph
  {et~al.}(2024)\citenamefont {{Mu{\~n}oz de las Heras}}, \citenamefont
  {Tabares}, \citenamefont {Schneider}, \citenamefont {Tagliacozzo},
  \citenamefont {Porras},\ and\ \citenamefont
  {{Gonz{\'a}lez-Tudela}}}]{munozdelasheras2024}%
  \BibitemOpen
  \bibfield  {author} {\bibinfo {author} {\bibfnamefont {A.}~\bibnamefont
  {{Mu{\~n}oz de las Heras}}}, \bibinfo {author} {\bibfnamefont
  {C.}~\bibnamefont {Tabares}}, \bibinfo {author} {\bibfnamefont {J.~T.}\
  \bibnamefont {Schneider}}, \bibinfo {author} {\bibfnamefont {L.}~\bibnamefont
  {Tagliacozzo}}, \bibinfo {author} {\bibfnamefont {D.}~\bibnamefont
  {Porras}},\ and\ \bibinfo {author} {\bibfnamefont {A.}~\bibnamefont
  {{Gonz{\'a}lez-Tudela}}},\ }\bibfield  {title} {\bibinfo {title} {Photonic
  quantum metrology with variational quantum optical nonlinearities},\ }\href
  {https://doi.org/10.1103/PhysRevResearch.6.013299} {\bibfield  {journal}
  {\bibinfo  {journal} {Phys. Rev. Res.}\ }\textbf {\bibinfo {volume} {6}},\
  \bibinfo {pages} {013299} (\bibinfo {year} {2024})}\BibitemShut {NoStop}%
\bibitem [{\citenamefont {Zanardi}\ \emph {et~al.}(2008)\citenamefont
  {Zanardi}, \citenamefont {Paris},\ and\ \citenamefont
  {Campos~Venuti}}]{zanardi2008}%
  \BibitemOpen
  \bibfield  {author} {\bibinfo {author} {\bibfnamefont {P.}~\bibnamefont
  {Zanardi}}, \bibinfo {author} {\bibfnamefont {M.~G.~A.}\ \bibnamefont
  {Paris}},\ and\ \bibinfo {author} {\bibfnamefont {L.}~\bibnamefont
  {Campos~Venuti}},\ }\bibfield  {title} {\bibinfo {title} {Quantum criticality
  as a resource for quantum estimation},\ }\href
  {https://doi.org/10.1103/PhysRevA.78.042105} {\bibfield  {journal} {\bibinfo
  {journal} {Phys. Rev. A}\ }\textbf {\bibinfo {volume} {78}},\ \bibinfo
  {pages} {042105} (\bibinfo {year} {2008})}\BibitemShut {NoStop}%
\bibitem [{\citenamefont {Fr{\'e}rot}\ and\ \citenamefont
  {Roscilde}(2018)}]{frerot2018}%
  \BibitemOpen
  \bibfield  {author} {\bibinfo {author} {\bibfnamefont {I.}~\bibnamefont
  {Fr{\'e}rot}}\ and\ \bibinfo {author} {\bibfnamefont {T.}~\bibnamefont
  {Roscilde}},\ }\bibfield  {title} {\bibinfo {title} {Quantum critical
  metrology},\ }\href {https://doi.org/10.1103/PhysRevLett.121.020402}
  {\bibfield  {journal} {\bibinfo  {journal} {Phys. Rev. Lett.}\ }\textbf
  {\bibinfo {volume} {121}},\ \bibinfo {pages} {020402} (\bibinfo {year}
  {2018})}\BibitemShut {NoStop}%
\bibitem [{\citenamefont {Chu}\ \emph {et~al.}(2021)\citenamefont {Chu},
  \citenamefont {Zhang}, \citenamefont {Yu},\ and\ \citenamefont
  {Cai}}]{chu2021}%
  \BibitemOpen
  \bibfield  {author} {\bibinfo {author} {\bibfnamefont {Y.}~\bibnamefont
  {Chu}}, \bibinfo {author} {\bibfnamefont {S.}~\bibnamefont {Zhang}}, \bibinfo
  {author} {\bibfnamefont {B.}~\bibnamefont {Yu}},\ and\ \bibinfo {author}
  {\bibfnamefont {J.}~\bibnamefont {Cai}},\ }\bibfield  {title} {\bibinfo
  {title} {Dynamic framework for criticality-enhanced quantum sensing},\ }\href
  {https://doi.org/10.1103/PhysRevLett.126.010502} {\bibfield  {journal}
  {\bibinfo  {journal} {Phys. Rev. Lett.}\ }\textbf {\bibinfo {volume} {126}},\
  \bibinfo {pages} {010502} (\bibinfo {year} {2021})}\BibitemShut {NoStop}%
\bibitem [{\citenamefont {Ding}\ \emph {et~al.}(2022)\citenamefont {Ding},
  \citenamefont {Liu}, \citenamefont {Shi}, \citenamefont {Guo}, \citenamefont
  {M{\o}lmer},\ and\ \citenamefont {Adams}}]{ding2022}%
  \BibitemOpen
  \bibfield  {author} {\bibinfo {author} {\bibfnamefont {D.-S.}\ \bibnamefont
  {Ding}}, \bibinfo {author} {\bibfnamefont {Z.-K.}\ \bibnamefont {Liu}},
  \bibinfo {author} {\bibfnamefont {B.-S.}\ \bibnamefont {Shi}}, \bibinfo
  {author} {\bibfnamefont {G.-C.}\ \bibnamefont {Guo}}, \bibinfo {author}
  {\bibfnamefont {K.}~\bibnamefont {M{\o}lmer}},\ and\ \bibinfo {author}
  {\bibfnamefont {C.~S.}\ \bibnamefont {Adams}},\ }\bibfield  {title} {\bibinfo
  {title} {Enhanced metrology at the critical point of a many-body rydberg
  atomic system},\ }\href {https://doi.org/10.1038/s41567-022-01777-8}
  {\bibfield  {journal} {\bibinfo  {journal} {Nat. Phys.}\ }\textbf {\bibinfo
  {volume} {18}},\ \bibinfo {pages} {1447} (\bibinfo {year}
  {2022})}\BibitemShut {NoStop}%
\bibitem [{\citenamefont {Schneider}\ \emph {et~al.}(1972)\citenamefont
  {Schneider}, \citenamefont {Stoll},\ and\ \citenamefont
  {Binder}}]{schneider1972}%
  \BibitemOpen
  \bibfield  {author} {\bibinfo {author} {\bibfnamefont {T.}~\bibnamefont
  {Schneider}}, \bibinfo {author} {\bibfnamefont {E.}~\bibnamefont {Stoll}},\
  and\ \bibinfo {author} {\bibfnamefont {K.}~\bibnamefont {Binder}},\
  }\bibfield  {title} {\bibinfo {title} {Critical slowing down in the kinetic
  ising model; evidence for the failure of the dynamical scaling hypothesis},\
  }\href {https://doi.org/10.1103/PhysRevLett.29.1080} {\bibfield  {journal}
  {\bibinfo  {journal} {Phys. Rev. Lett.}\ }\textbf {\bibinfo {volume} {29}},\
  \bibinfo {pages} {1080} (\bibinfo {year} {1972})}\BibitemShut {NoStop}%
\bibitem [{\citenamefont {Brookes}\ \emph {et~al.}(2021)\citenamefont
  {Brookes}, \citenamefont {Tancredi}, \citenamefont {Patterson}, \citenamefont
  {Rahamim}, \citenamefont {Esposito}, \citenamefont {Mavrogordatos},
  \citenamefont {Leek}, \citenamefont {Ginossar},\ and\ \citenamefont
  {Szymanska}}]{brookes2021}%
  \BibitemOpen
  \bibfield  {author} {\bibinfo {author} {\bibfnamefont {P.}~\bibnamefont
  {Brookes}}, \bibinfo {author} {\bibfnamefont {G.}~\bibnamefont {Tancredi}},
  \bibinfo {author} {\bibfnamefont {A.~D.}\ \bibnamefont {Patterson}}, \bibinfo
  {author} {\bibfnamefont {J.}~\bibnamefont {Rahamim}}, \bibinfo {author}
  {\bibfnamefont {M.}~\bibnamefont {Esposito}}, \bibinfo {author}
  {\bibfnamefont {T.~K.}\ \bibnamefont {Mavrogordatos}}, \bibinfo {author}
  {\bibfnamefont {P.~J.}\ \bibnamefont {Leek}}, \bibinfo {author}
  {\bibfnamefont {E.}~\bibnamefont {Ginossar}},\ and\ \bibinfo {author}
  {\bibfnamefont {M.~H.}\ \bibnamefont {Szymanska}},\ }\bibfield  {title}
  {\bibinfo {title} {Critical slowing down in circuit quantum
  electrodynamics},\ }\href {https://doi.org/10.1126/sciadv.abe9492} {\bibfield
   {journal} {\bibinfo  {journal} {Sci. Adv.}\ }\textbf {\bibinfo {volume}
  {7}},\ \bibinfo {pages} {eabe9492} (\bibinfo {year} {2021})}\BibitemShut
  {NoStop}%
\bibitem [{\citenamefont {Luis}(2004)}]{luis2004}%
  \BibitemOpen
  \bibfield  {author} {\bibinfo {author} {\bibfnamefont {A.}~\bibnamefont
  {Luis}},\ }\bibfield  {title} {\bibinfo {title} {Nonlinear transformations
  and the heisenberg limit},\ }\href
  {https://doi.org/10.1016/j.physleta.2004.06.080} {\bibfield  {journal}
  {\bibinfo  {journal} {Phys. Lett. A}\ }\textbf {\bibinfo {volume} {329}},\
  \bibinfo {pages} {8} (\bibinfo {year} {2004})}\BibitemShut {NoStop}%
\bibitem [{\citenamefont {Boixo}\ \emph {et~al.}(2008)\citenamefont {Boixo},
  \citenamefont {Datta}, \citenamefont {Davis}, \citenamefont {Flammia},
  \citenamefont {Shaji},\ and\ \citenamefont {Caves}}]{boixo2008}%
  \BibitemOpen
  \bibfield  {author} {\bibinfo {author} {\bibfnamefont {S.}~\bibnamefont
  {Boixo}}, \bibinfo {author} {\bibfnamefont {A.}~\bibnamefont {Datta}},
  \bibinfo {author} {\bibfnamefont {M.~J.}\ \bibnamefont {Davis}}, \bibinfo
  {author} {\bibfnamefont {S.~T.}\ \bibnamefont {Flammia}}, \bibinfo {author}
  {\bibfnamefont {A.}~\bibnamefont {Shaji}},\ and\ \bibinfo {author}
  {\bibfnamefont {C.~M.}\ \bibnamefont {Caves}},\ }\bibfield  {title} {\bibinfo
  {title} {Quantum metrology: Dynamics versus entanglement},\ }\href
  {https://doi.org/10.1103/PhysRevLett.101.040403} {\bibfield  {journal}
  {\bibinfo  {journal} {Phys. Rev. Lett.}\ }\textbf {\bibinfo {volume} {101}},\
  \bibinfo {pages} {040403} (\bibinfo {year} {2008})}\BibitemShut {NoStop}%
\bibitem [{\citenamefont {Boixo}\ \emph {et~al.}(2007)\citenamefont {Boixo},
  \citenamefont {Flammia}, \citenamefont {Caves},\ and\ \citenamefont
  {Geremia}}]{boixo2007}%
  \BibitemOpen
  \bibfield  {author} {\bibinfo {author} {\bibfnamefont {S.}~\bibnamefont
  {Boixo}}, \bibinfo {author} {\bibfnamefont {S.~T.}\ \bibnamefont {Flammia}},
  \bibinfo {author} {\bibfnamefont {C.~M.}\ \bibnamefont {Caves}},\ and\
  \bibinfo {author} {\bibfnamefont {{\relax JM}.}~\bibnamefont {Geremia}},\
  }\bibfield  {title} {\bibinfo {title} {Generalized limits for
  single-parameter quantum estimation},\ }\href
  {https://doi.org/10.1103/PhysRevLett.98.090401} {\bibfield  {journal}
  {\bibinfo  {journal} {Phys. Rev. Lett.}\ }\textbf {\bibinfo {volume} {98}},\
  \bibinfo {pages} {090401} (\bibinfo {year} {2007})}\BibitemShut {NoStop}%
\bibitem [{\citenamefont {Napolitano}\ \emph {et~al.}(2011)\citenamefont
  {Napolitano}, \citenamefont {Koschorreck}, \citenamefont {Dubost},
  \citenamefont {Behbood}, \citenamefont {Sewell},\ and\ \citenamefont
  {Mitchell}}]{napolitano2011}%
  \BibitemOpen
  \bibfield  {author} {\bibinfo {author} {\bibfnamefont {M.}~\bibnamefont
  {Napolitano}}, \bibinfo {author} {\bibfnamefont {M.}~\bibnamefont
  {Koschorreck}}, \bibinfo {author} {\bibfnamefont {B.}~\bibnamefont {Dubost}},
  \bibinfo {author} {\bibfnamefont {N.}~\bibnamefont {Behbood}}, \bibinfo
  {author} {\bibfnamefont {R.~J.}\ \bibnamefont {Sewell}},\ and\ \bibinfo
  {author} {\bibfnamefont {M.~W.}\ \bibnamefont {Mitchell}},\ }\bibfield
  {title} {\bibinfo {title} {Interaction-based quantum metrology showing
  scaling beyond the heisenberg limit},\ }\href
  {https://doi.org/10.1038/nature09778} {\bibfield  {journal} {\bibinfo
  {journal} {Nature}\ }\textbf {\bibinfo {volume} {471}},\ \bibinfo {pages}
  {486} (\bibinfo {year} {2011})}\BibitemShut {NoStop}%
\bibitem [{\citenamefont {Montenegro}\ \emph {et~al.}(2025)\citenamefont
  {Montenegro}, \citenamefont {Mukhopadhyay}, \citenamefont {Yousefjani},
  \citenamefont {Sarkar}, \citenamefont {Mishra}, \citenamefont {Paris},\ and\
  \citenamefont {Bayat}}]{Montenegro2025}%
  \BibitemOpen
  \bibfield  {author} {\bibinfo {author} {\bibfnamefont {V.}~\bibnamefont
  {Montenegro}}, \bibinfo {author} {\bibfnamefont {C.}~\bibnamefont
  {Mukhopadhyay}}, \bibinfo {author} {\bibfnamefont {R.}~\bibnamefont
  {Yousefjani}}, \bibinfo {author} {\bibfnamefont {S.}~\bibnamefont {Sarkar}},
  \bibinfo {author} {\bibfnamefont {U.}~\bibnamefont {Mishra}}, \bibinfo
  {author} {\bibfnamefont {M.~G.}\ \bibnamefont {Paris}},\ and\ \bibinfo
  {author} {\bibfnamefont {A.}~\bibnamefont {Bayat}},\ }\bibfield  {title}
  {\bibinfo {title} {Review: Quantum metrology and sensing with many-body
  systems},\ }\href
  {https://doi.org/https://doi.org/10.1016/j.physrep.2025.05.005} {\bibfield
  {journal} {\bibinfo  {journal} {Physics Reports}\ }\textbf {\bibinfo {volume}
  {1134}},\ \bibinfo {pages} {1} (\bibinfo {year} {2025})},\ \bibinfo {note}
  {review: Quantum metrology and sensing with many-body systems}\BibitemShut
  {NoStop}%
\bibitem [{\citenamefont {Domokos}\ \emph {et~al.}(2002)\citenamefont
  {Domokos}, \citenamefont {Horak},\ and\ \citenamefont
  {Ritsch}}]{domokos2002}%
  \BibitemOpen
  \bibfield  {author} {\bibinfo {author} {\bibfnamefont {P.}~\bibnamefont
  {Domokos}}, \bibinfo {author} {\bibfnamefont {P.}~\bibnamefont {Horak}},\
  and\ \bibinfo {author} {\bibfnamefont {H.}~\bibnamefont {Ritsch}},\
  }\bibfield  {title} {\bibinfo {title} {Quantum description of light-pulse
  scattering on a single atom in waveguides},\ }\href
  {https://doi.org/10.1103/PhysRevA.65.033832} {\bibfield  {journal} {\bibinfo
  {journal} {Phys. Rev. A}\ }\textbf {\bibinfo {volume} {65}},\ \bibinfo
  {pages} {033832} (\bibinfo {year} {2002})}\BibitemShut {NoStop}%
\bibitem [{\citenamefont {Shen}\ and\ \citenamefont {Fan}(2005)}]{shen2005a}%
  \BibitemOpen
  \bibfield  {author} {\bibinfo {author} {\bibfnamefont {J.-T.}\ \bibnamefont
  {Shen}}\ and\ \bibinfo {author} {\bibfnamefont {S.}~\bibnamefont {Fan}},\
  }\bibfield  {title} {\bibinfo {title} {Coherent single photon transport in a
  one-dimensional waveguide coupled with superconducting quantum bits},\ }\href
  {https://doi.org/10.1103/PhysRevLett.95.213001} {\bibfield  {journal}
  {\bibinfo  {journal} {Phys. Rev. Lett.}\ }\textbf {\bibinfo {volume} {95}},\
  \bibinfo {pages} {213001} (\bibinfo {year} {2005})}\BibitemShut {NoStop}%
\bibitem [{\citenamefont {Liao}\ and\ \citenamefont {Law}(2010)}]{liao2010}%
  \BibitemOpen
  \bibfield  {author} {\bibinfo {author} {\bibfnamefont {J.-Q.}\ \bibnamefont
  {Liao}}\ and\ \bibinfo {author} {\bibfnamefont {C.~K.}\ \bibnamefont {Law}},\
  }\bibfield  {title} {\bibinfo {title} {Correlated two-photon transport in a
  one-dimensional waveguide side-coupled to a nonlinear cavity},\ }\href
  {https://doi.org/10.1103/PhysRevA.82.053836} {\bibfield  {journal} {\bibinfo
  {journal} {Phys. Rev. A}\ }\textbf {\bibinfo {volume} {82}},\ \bibinfo
  {pages} {053836} (\bibinfo {year} {2010})}\BibitemShut {NoStop}%
\bibitem [{\citenamefont {Liao}\ \emph
  {et~al.}(2016{\natexlab{a}})\citenamefont {Liao}, \citenamefont {Zeng},
  \citenamefont {Nha},\ and\ \citenamefont {Zubairy}}]{liao2016a}%
  \BibitemOpen
  \bibfield  {author} {\bibinfo {author} {\bibfnamefont {Z.}~\bibnamefont
  {Liao}}, \bibinfo {author} {\bibfnamefont {X.}~\bibnamefont {Zeng}}, \bibinfo
  {author} {\bibfnamefont {H.}~\bibnamefont {Nha}},\ and\ \bibinfo {author}
  {\bibfnamefont {M.~S.}\ \bibnamefont {Zubairy}},\ }\bibfield  {title}
  {\bibinfo {title} {Photon transport in a one-dimensional nanophotonic
  waveguide {QED} system},\ }\href
  {https://doi.org/10.1088/0031-8949/91/6/063004} {\bibfield  {journal}
  {\bibinfo  {journal} {Phys. Scr.}\ }\textbf {\bibinfo {volume} {91}},\
  \bibinfo {pages} {063004} (\bibinfo {year} {2016}{\natexlab{a}})}\BibitemShut
  {NoStop}%
\bibitem [{\citenamefont {Sheremet}\ \emph {et~al.}(2023)\citenamefont
  {Sheremet}, \citenamefont {Petrov}, \citenamefont {Iorsh}, \citenamefont
  {Poshakinskiy},\ and\ \citenamefont {Poddubny}}]{sheremet2023}%
  \BibitemOpen
  \bibfield  {author} {\bibinfo {author} {\bibfnamefont {A.~S.}\ \bibnamefont
  {Sheremet}}, \bibinfo {author} {\bibfnamefont {M.~I.}\ \bibnamefont
  {Petrov}}, \bibinfo {author} {\bibfnamefont {I.~V.}\ \bibnamefont {Iorsh}},
  \bibinfo {author} {\bibfnamefont {A.~V.}\ \bibnamefont {Poshakinskiy}},\ and\
  \bibinfo {author} {\bibfnamefont {A.~N.}\ \bibnamefont {Poddubny}},\
  }\bibfield  {title} {\bibinfo {title} {Waveguide quantum electrodynamics:
  Collective radiance and photon-photon correlations},\ }\href
  {https://doi.org/10.1103/RevModPhys.95.015002} {\bibfield  {journal}
  {\bibinfo  {journal} {Rev. Mod. Phys.}\ }\textbf {\bibinfo {volume} {95}},\
  \bibinfo {pages} {015002} (\bibinfo {year} {2023})}\BibitemShut {NoStop}%
\bibitem [{\citenamefont {Roy}\ \emph {et~al.}(2017)\citenamefont {Roy},
  \citenamefont {Wilson},\ and\ \citenamefont {Firstenberg}}]{roy2017}%
  \BibitemOpen
  \bibfield  {author} {\bibinfo {author} {\bibfnamefont {D.}~\bibnamefont
  {Roy}}, \bibinfo {author} {\bibfnamefont {C.~M.}\ \bibnamefont {Wilson}},\
  and\ \bibinfo {author} {\bibfnamefont {O.}~\bibnamefont {Firstenberg}},\
  }\bibfield  {title} {\bibinfo {title} {Colloquium: Strongly interacting
  photons in one-dimensional continuum},\ }\href
  {https://doi.org/10.1103/RevModPhys.89.021001} {\bibfield  {journal}
  {\bibinfo  {journal} {Rev. Mod. Phys.}\ }\textbf {\bibinfo {volume} {89}},\
  \bibinfo {pages} {021001} (\bibinfo {year} {2017})}\BibitemShut {NoStop}%
\bibitem [{\citenamefont {Tian}\ \emph {et~al.}(2025)\citenamefont {Tian},
  \citenamefont {Zheng}, \citenamefont {Zhan}, \citenamefont {Nori},\ and\
  \citenamefont {L{\"u}}}]{tian2025}%
  \BibitemOpen
  \bibfield  {author} {\bibinfo {author} {\bibfnamefont {G.}~\bibnamefont
  {Tian}}, \bibinfo {author} {\bibfnamefont {L.-L.}\ \bibnamefont {Zheng}},
  \bibinfo {author} {\bibfnamefont {Z.-M.}\ \bibnamefont {Zhan}}, \bibinfo
  {author} {\bibfnamefont {F.}~\bibnamefont {Nori}},\ and\ \bibinfo {author}
  {\bibfnamefont {X.-Y.}\ \bibnamefont {L{\"u}}},\ }\bibfield  {title}
  {\bibinfo {title} {Disorder-induced strongly correlated photons in waveguide
  {QED}},\ }\href {https://doi.org/10.1103/mldt-d59t} {\bibfield  {journal}
  {\bibinfo  {journal} {Phys. Rev. Lett.}\ }\textbf {\bibinfo {volume} {135}},\
  \bibinfo {pages} {153604} (\bibinfo {year} {2025})}\BibitemShut {NoStop}%
\bibitem [{\citenamefont {Liao}\ \emph {et~al.}(2015)\citenamefont {Liao},
  \citenamefont {Zeng}, \citenamefont {Zhu},\ and\ \citenamefont
  {Zubairy}}]{zeyangliao2015}%
  \BibitemOpen
  \bibfield  {author} {\bibinfo {author} {\bibfnamefont {Z.}~\bibnamefont
  {Liao}}, \bibinfo {author} {\bibfnamefont {X.}~\bibnamefont {Zeng}}, \bibinfo
  {author} {\bibfnamefont {S.-Y.}\ \bibnamefont {Zhu}},\ and\ \bibinfo {author}
  {\bibfnamefont {M.~S.}\ \bibnamefont {Zubairy}},\ }\bibfield  {title}
  {\bibinfo {title} {Single-photon transport through an atomic chain coupled to
  a one-dimensional nanophotonic waveguide},\ }\href
  {https://doi.org/10.1103/PhysRevA.92.023806} {\bibfield  {journal} {\bibinfo
  {journal} {Phys. Rev. A}\ }\textbf {\bibinfo {volume} {92}},\ \bibinfo
  {pages} {023806} (\bibinfo {year} {2015})}\BibitemShut {NoStop}%
\bibitem [{\citenamefont {Goban}\ \emph {et~al.}(2015)\citenamefont {Goban},
  \citenamefont {Hung}, \citenamefont {Hood}, \citenamefont {Yu}, \citenamefont
  {Muniz}, \citenamefont {Painter},\ and\ \citenamefont {Kimble}}]{goban2015}%
  \BibitemOpen
  \bibfield  {author} {\bibinfo {author} {\bibfnamefont {A.}~\bibnamefont
  {Goban}}, \bibinfo {author} {\bibfnamefont {C.-L.}\ \bibnamefont {Hung}},
  \bibinfo {author} {\bibfnamefont {J.~D.}\ \bibnamefont {Hood}}, \bibinfo
  {author} {\bibfnamefont {S.-P.}\ \bibnamefont {Yu}}, \bibinfo {author}
  {\bibfnamefont {J.~A.}\ \bibnamefont {Muniz}}, \bibinfo {author}
  {\bibfnamefont {O.}~\bibnamefont {Painter}},\ and\ \bibinfo {author}
  {\bibfnamefont {H.~J.}\ \bibnamefont {Kimble}},\ }\bibfield  {title}
  {\bibinfo {title} {Superradiance for atoms trapped along a photonic crystal
  waveguide},\ }\href {https://doi.org/10.1103/PhysRevLett.115.063601}
  {\bibfield  {journal} {\bibinfo  {journal} {Phys. Rev. Lett.}\ }\textbf
  {\bibinfo {volume} {115}},\ \bibinfo {pages} {063601} (\bibinfo {year}
  {2015})}\BibitemShut {NoStop}%
\bibitem [{\citenamefont {Perczel}\ \emph {et~al.}(2017)\citenamefont
  {Perczel}, \citenamefont {Borregaard}, \citenamefont {Chang}, \citenamefont
  {Pichler}, \citenamefont {Yelin}, \citenamefont {Zoller},\ and\ \citenamefont
  {Lukin}}]{perczel2017}%
  \BibitemOpen
  \bibfield  {author} {\bibinfo {author} {\bibfnamefont {J.}~\bibnamefont
  {Perczel}}, \bibinfo {author} {\bibfnamefont {J.}~\bibnamefont {Borregaard}},
  \bibinfo {author} {\bibfnamefont {D.~E.}\ \bibnamefont {Chang}}, \bibinfo
  {author} {\bibfnamefont {H.}~\bibnamefont {Pichler}}, \bibinfo {author}
  {\bibfnamefont {S.~F.}\ \bibnamefont {Yelin}}, \bibinfo {author}
  {\bibfnamefont {P.}~\bibnamefont {Zoller}},\ and\ \bibinfo {author}
  {\bibfnamefont {M.~D.}\ \bibnamefont {Lukin}},\ }\bibfield  {title} {\bibinfo
  {title} {Topological quantum optics in two-dimensional atomic arrays},\
  }\href {https://doi.org/10.1103/PhysRevLett.119.023603} {\bibfield  {journal}
  {\bibinfo  {journal} {Phys. Rev. Lett.}\ }\textbf {\bibinfo {volume} {119}},\
  \bibinfo {pages} {023603} (\bibinfo {year} {2017})}\BibitemShut {NoStop}%
\bibitem [{\citenamefont {Cheng}\ \emph {et~al.}(2017)\citenamefont {Cheng},
  \citenamefont {Xu},\ and\ \citenamefont {Agarwal}}]{cheng2017}%
  \BibitemOpen
  \bibfield  {author} {\bibinfo {author} {\bibfnamefont {M.-T.}\ \bibnamefont
  {Cheng}}, \bibinfo {author} {\bibfnamefont {J.}~\bibnamefont {Xu}},\ and\
  \bibinfo {author} {\bibfnamefont {G.~S.}\ \bibnamefont {Agarwal}},\
  }\bibfield  {title} {\bibinfo {title} {Waveguide transport mediated by strong
  coupling with atoms},\ }\href {https://doi.org/10.1103/PhysRevA.95.053807}
  {\bibfield  {journal} {\bibinfo  {journal} {Phys. Rev. A}\ }\textbf {\bibinfo
  {volume} {95}},\ \bibinfo {pages} {053807} (\bibinfo {year}
  {2017})}\BibitemShut {NoStop}%
\bibitem [{\citenamefont {Song}\ \emph {et~al.}(2017)\citenamefont {Song},
  \citenamefont {Munro}, \citenamefont {Nie}, \citenamefont {Deng},
  \citenamefont {Yang},\ and\ \citenamefont {Kwek}}]{song2017}%
  \BibitemOpen
  \bibfield  {author} {\bibinfo {author} {\bibfnamefont {G.-Z.}\ \bibnamefont
  {Song}}, \bibinfo {author} {\bibfnamefont {E.}~\bibnamefont {Munro}},
  \bibinfo {author} {\bibfnamefont {W.}~\bibnamefont {Nie}}, \bibinfo {author}
  {\bibfnamefont {F.-G.}\ \bibnamefont {Deng}}, \bibinfo {author}
  {\bibfnamefont {G.-J.}\ \bibnamefont {Yang}},\ and\ \bibinfo {author}
  {\bibfnamefont {L.-C.}\ \bibnamefont {Kwek}},\ }\bibfield  {title} {\bibinfo
  {title} {Photon scattering by an atomic ensemble coupled to a one-dimensional
  nanophotonic waveguide},\ }\href {https://doi.org/10.1103/PhysRevA.96.043872}
  {\bibfield  {journal} {\bibinfo  {journal} {Phys. Rev. A}\ }\textbf {\bibinfo
  {volume} {96}},\ \bibinfo {pages} {043872} (\bibinfo {year}
  {2017})}\BibitemShut {NoStop}%
\bibitem [{\citenamefont {Chang}\ \emph {et~al.}(2018)\citenamefont {Chang},
  \citenamefont {Douglas}, \citenamefont {{Gonz{\'a}lez-Tudela}}, \citenamefont
  {Hung},\ and\ \citenamefont {Kimble}}]{chang2018a}%
  \BibitemOpen
  \bibfield  {author} {\bibinfo {author} {\bibfnamefont {D.~E.}\ \bibnamefont
  {Chang}}, \bibinfo {author} {\bibfnamefont {J.~S.}\ \bibnamefont {Douglas}},
  \bibinfo {author} {\bibfnamefont {A.}~\bibnamefont {{Gonz{\'a}lez-Tudela}}},
  \bibinfo {author} {\bibfnamefont {C.-L.}\ \bibnamefont {Hung}},\ and\
  \bibinfo {author} {\bibfnamefont {H.~J.}\ \bibnamefont {Kimble}},\ }\bibfield
   {title} {\bibinfo {title} {Colloquium: Quantum matter built from nanoscopic
  lattices of atoms and photons},\ }\href
  {https://doi.org/10.1103/RevModPhys.90.031002} {\bibfield  {journal}
  {\bibinfo  {journal} {Rev. Mod. Phys.}\ }\textbf {\bibinfo {volume} {90}},\
  \bibinfo {pages} {031002} (\bibinfo {year} {2018})}\BibitemShut {NoStop}%
\bibitem [{\citenamefont {Fayard}\ \emph {et~al.}(2021)\citenamefont {Fayard},
  \citenamefont {Henriet}, \citenamefont {{Asenjo-Garcia}},\ and\ \citenamefont
  {Chang}}]{fayard2021a}%
  \BibitemOpen
  \bibfield  {author} {\bibinfo {author} {\bibfnamefont {N.}~\bibnamefont
  {Fayard}}, \bibinfo {author} {\bibfnamefont {L.}~\bibnamefont {Henriet}},
  \bibinfo {author} {\bibfnamefont {A.}~\bibnamefont {{Asenjo-Garcia}}},\ and\
  \bibinfo {author} {\bibfnamefont {D.~E.}\ \bibnamefont {Chang}},\ }\bibfield
  {title} {\bibinfo {title} {Many-body localization in waveguide quantum
  electrodynamics},\ }\href {https://doi.org/10.1103/PhysRevResearch.3.033233}
  {\bibfield  {journal} {\bibinfo  {journal} {Phys. Rev. Res.}\ }\textbf
  {\bibinfo {volume} {3}},\ \bibinfo {pages} {033233} (\bibinfo {year}
  {2021})}\BibitemShut {NoStop}%
\bibitem [{\citenamefont {Reitz}\ \emph {et~al.}(2022)\citenamefont {Reitz},
  \citenamefont {Sommer},\ and\ \citenamefont {Genes}}]{reitz2022}%
  \BibitemOpen
  \bibfield  {author} {\bibinfo {author} {\bibfnamefont {M.}~\bibnamefont
  {Reitz}}, \bibinfo {author} {\bibfnamefont {C.}~\bibnamefont {Sommer}},\ and\
  \bibinfo {author} {\bibfnamefont {C.}~\bibnamefont {Genes}},\ }\bibfield
  {title} {\bibinfo {title} {Cooperative quantum phenomena in light-matter
  platforms},\ }\href {https://doi.org/10.1103/PRXQuantum.3.010201} {\bibfield
  {journal} {\bibinfo  {journal} {PRX Quantum}\ }\textbf {\bibinfo {volume}
  {3}},\ \bibinfo {pages} {010201} (\bibinfo {year} {2022})}\BibitemShut
  {NoStop}%
\bibitem [{\citenamefont {Nie}\ \emph {et~al.}(2023)\citenamefont {Nie},
  \citenamefont {Shi}, \citenamefont {Liu},\ and\ \citenamefont
  {Nori}}]{nie2023}%
  \BibitemOpen
  \bibfield  {author} {\bibinfo {author} {\bibfnamefont {W.}~\bibnamefont
  {Nie}}, \bibinfo {author} {\bibfnamefont {T.}~\bibnamefont {Shi}}, \bibinfo
  {author} {\bibfnamefont {Y.-x.}\ \bibnamefont {Liu}},\ and\ \bibinfo {author}
  {\bibfnamefont {F.}~\bibnamefont {Nori}},\ }\bibfield  {title} {\bibinfo
  {title} {Non-hermitian waveguide cavity {QED} with tunable atomic mirrors},\
  }\href {https://doi.org/10.1103/PhysRevLett.131.103602} {\bibfield  {journal}
  {\bibinfo  {journal} {Phys. Rev. Lett.}\ }\textbf {\bibinfo {volume} {131}},\
  \bibinfo {pages} {103602} (\bibinfo {year} {2023})}\BibitemShut {NoStop}%
\bibitem [{\citenamefont {Lu}\ \emph {et~al.}(2024)\citenamefont {Lu},
  \citenamefont {Liu}, \citenamefont {Jiang},\ and\ \citenamefont
  {Liao}}]{lu2024}%
  \BibitemOpen
  \bibfield  {author} {\bibinfo {author} {\bibfnamefont {Y.-W.}\ \bibnamefont
  {Lu}}, \bibinfo {author} {\bibfnamefont {J.-F.}\ \bibnamefont {Liu}},
  \bibinfo {author} {\bibfnamefont {H.}~\bibnamefont {Jiang}},\ and\ \bibinfo
  {author} {\bibfnamefont {Z.}~\bibnamefont {Liao}},\ }\bibfield  {title}
  {\bibinfo {title} {Topologically protected subradiant cavity polaritons
  through linewidth narrowing enabled by dissipationless edge states},\ }\href
  {https://doi.org/10.1088/2058-9565/ad3f46} {\bibfield  {journal} {\bibinfo
  {journal} {Quantum Sci. Technol.}\ }\textbf {\bibinfo {volume} {9}},\
  \bibinfo {pages} {035019} (\bibinfo {year} {2024})}\BibitemShut {NoStop}%
\bibitem [{\citenamefont {Paulisch}\ \emph {et~al.}(2016)\citenamefont
  {Paulisch}, \citenamefont {Kimble},\ and\ \citenamefont
  {{Gonz{\'a}lez-Tudela}}}]{paulisch2016}%
  \BibitemOpen
  \bibfield  {author} {\bibinfo {author} {\bibfnamefont {V.}~\bibnamefont
  {Paulisch}}, \bibinfo {author} {\bibfnamefont {H.~J.}\ \bibnamefont
  {Kimble}},\ and\ \bibinfo {author} {\bibfnamefont {A.}~\bibnamefont
  {{Gonz{\'a}lez-Tudela}}},\ }\bibfield  {title} {\bibinfo {title} {Universal
  quantum computation in waveguide {QED} using decoherence free subspaces},\
  }\href {https://doi.org/10.1088/1367-2630/18/4/043041} {\bibfield  {journal}
  {\bibinfo  {journal} {New J. Phys.}\ }\textbf {\bibinfo {volume} {18}},\
  \bibinfo {pages} {043041} (\bibinfo {year} {2016})}\BibitemShut {NoStop}%
\bibitem [{\citenamefont {Albrecht}\ \emph {et~al.}(2019)\citenamefont
  {Albrecht}, \citenamefont {Henriet}, \citenamefont {{Asenjo-Garcia}},
  \citenamefont {Dieterle}, \citenamefont {Painter},\ and\ \citenamefont
  {Chang}}]{albrecht2019}%
  \BibitemOpen
  \bibfield  {author} {\bibinfo {author} {\bibfnamefont {A.}~\bibnamefont
  {Albrecht}}, \bibinfo {author} {\bibfnamefont {L.}~\bibnamefont {Henriet}},
  \bibinfo {author} {\bibfnamefont {A.}~\bibnamefont {{Asenjo-Garcia}}},
  \bibinfo {author} {\bibfnamefont {P.~B.}\ \bibnamefont {Dieterle}}, \bibinfo
  {author} {\bibfnamefont {O.}~\bibnamefont {Painter}},\ and\ \bibinfo {author}
  {\bibfnamefont {D.~E.}\ \bibnamefont {Chang}},\ }\bibfield  {title} {\bibinfo
  {title} {Subradiant states of quantum bits coupled to a one-dimensional
  waveguide},\ }\href {https://doi.org/10.1088/1367-2630/ab0134} {\bibfield
  {journal} {\bibinfo  {journal} {New J. Phys.}\ }\textbf {\bibinfo {volume}
  {21}},\ \bibinfo {pages} {025003} (\bibinfo {year} {2019})}\BibitemShut
  {NoStop}%
\bibitem [{\citenamefont {Bluvstein}\ \emph {et~al.}(2022)\citenamefont
  {Bluvstein}, \citenamefont {Levine}, \citenamefont {Semeghini}, \citenamefont
  {Wang}, \citenamefont {Ebadi}, \citenamefont {Kalinowski}, \citenamefont
  {Keesling}, \citenamefont {Maskara}, \citenamefont {Pichler}, \citenamefont
  {Greiner}, \citenamefont {Vuleti{\'c}},\ and\ \citenamefont
  {Lukin}}]{bluvstein2022}%
  \BibitemOpen
  \bibfield  {author} {\bibinfo {author} {\bibfnamefont {D.}~\bibnamefont
  {Bluvstein}}, \bibinfo {author} {\bibfnamefont {H.}~\bibnamefont {Levine}},
  \bibinfo {author} {\bibfnamefont {G.}~\bibnamefont {Semeghini}}, \bibinfo
  {author} {\bibfnamefont {T.~T.}\ \bibnamefont {Wang}}, \bibinfo {author}
  {\bibfnamefont {S.}~\bibnamefont {Ebadi}}, \bibinfo {author} {\bibfnamefont
  {M.}~\bibnamefont {Kalinowski}}, \bibinfo {author} {\bibfnamefont
  {A.}~\bibnamefont {Keesling}}, \bibinfo {author} {\bibfnamefont
  {N.}~\bibnamefont {Maskara}}, \bibinfo {author} {\bibfnamefont
  {H.}~\bibnamefont {Pichler}}, \bibinfo {author} {\bibfnamefont
  {M.}~\bibnamefont {Greiner}}, \bibinfo {author} {\bibfnamefont
  {V.}~\bibnamefont {Vuleti{\'c}}},\ and\ \bibinfo {author} {\bibfnamefont
  {M.~D.}\ \bibnamefont {Lukin}},\ }\bibfield  {title} {\bibinfo {title} {A
  quantum processor based on coherent transport of entangled atom arrays},\
  }\href {https://doi.org/10.1038/s41586-022-04592-6} {\bibfield  {journal}
  {\bibinfo  {journal} {Nature}\ }\textbf {\bibinfo {volume} {604}},\ \bibinfo
  {pages} {451} (\bibinfo {year} {2022})}\BibitemShut {NoStop}%
\bibitem [{\citenamefont {Masson}\ and\ \citenamefont
  {{Asenjo-Garcia}}(2020)}]{masson2020}%
  \BibitemOpen
  \bibfield  {author} {\bibinfo {author} {\bibfnamefont {S.~J.}\ \bibnamefont
  {Masson}}\ and\ \bibinfo {author} {\bibfnamefont {A.}~\bibnamefont
  {{Asenjo-Garcia}}},\ }\bibfield  {title} {\bibinfo {title} {Atomic-waveguide
  quantum electrodynamics},\ }\href
  {https://doi.org/10.1103/PhysRevResearch.2.043213} {\bibfield  {journal}
  {\bibinfo  {journal} {Phys. Rev. Research}\ }\textbf {\bibinfo {volume}
  {2}},\ \bibinfo {pages} {043213} (\bibinfo {year} {2020})}\BibitemShut
  {NoStop}%
\bibitem [{\citenamefont {Zhang}\ and\ \citenamefont
  {M{\o}lmer}(2019)}]{zhang2019a}%
  \BibitemOpen
  \bibfield  {author} {\bibinfo {author} {\bibfnamefont {Y.-X.}\ \bibnamefont
  {Zhang}}\ and\ \bibinfo {author} {\bibfnamefont {K.}~\bibnamefont
  {M{\o}lmer}},\ }\bibfield  {title} {\bibinfo {title} {Theory of subradiant
  states of a one-dimensional two-level atom chain},\ }\href
  {https://doi.org/10.1103/PhysRevLett.122.203605} {\bibfield  {journal}
  {\bibinfo  {journal} {Phys. Rev. Lett.}\ }\textbf {\bibinfo {volume} {122}},\
  \bibinfo {pages} {203605} (\bibinfo {year} {2019})}\BibitemShut {NoStop}%
\bibitem [{\citenamefont {Shahmoon}\ \emph {et~al.}(2017)\citenamefont
  {Shahmoon}, \citenamefont {Wild}, \citenamefont {Lukin},\ and\ \citenamefont
  {Yelin}}]{shahmoon2017}%
  \BibitemOpen
  \bibfield  {author} {\bibinfo {author} {\bibfnamefont {E.}~\bibnamefont
  {Shahmoon}}, \bibinfo {author} {\bibfnamefont {D.~S.}\ \bibnamefont {Wild}},
  \bibinfo {author} {\bibfnamefont {M.~D.}\ \bibnamefont {Lukin}},\ and\
  \bibinfo {author} {\bibfnamefont {S.~F.}\ \bibnamefont {Yelin}},\ }\bibfield
  {title} {\bibinfo {title} {Cooperative resonances in light scattering from
  two-dimensional atomic arrays},\ }\href
  {https://doi.org/10.1103/PhysRevLett.118.113601} {\bibfield  {journal}
  {\bibinfo  {journal} {Phys. Rev. Lett.}\ }\textbf {\bibinfo {volume} {118}},\
  \bibinfo {pages} {113601} (\bibinfo {year} {2017})}\BibitemShut {NoStop}%
\bibitem [{\citenamefont {Rui}\ \emph {et~al.}(2020)\citenamefont {Rui},
  \citenamefont {Wei}, \citenamefont {{Rubio-Abadal}}, \citenamefont
  {Hollerith}, \citenamefont {Zeiher}, \citenamefont {{Stamper-Kurn}},
  \citenamefont {Gross},\ and\ \citenamefont {Bloch}}]{rui2020}%
  \BibitemOpen
  \bibfield  {author} {\bibinfo {author} {\bibfnamefont {J.}~\bibnamefont
  {Rui}}, \bibinfo {author} {\bibfnamefont {D.}~\bibnamefont {Wei}}, \bibinfo
  {author} {\bibfnamefont {A.}~\bibnamefont {{Rubio-Abadal}}}, \bibinfo
  {author} {\bibfnamefont {S.}~\bibnamefont {Hollerith}}, \bibinfo {author}
  {\bibfnamefont {J.}~\bibnamefont {Zeiher}}, \bibinfo {author} {\bibfnamefont
  {D.~M.}\ \bibnamefont {{Stamper-Kurn}}}, \bibinfo {author} {\bibfnamefont
  {C.}~\bibnamefont {Gross}},\ and\ \bibinfo {author} {\bibfnamefont
  {I.}~\bibnamefont {Bloch}},\ }\bibfield  {title} {\bibinfo {title} {A
  subradiant optical mirror formed by a single structured atomic layer},\
  }\href {https://doi.org/10.1038/s41586-020-2463-x} {\bibfield  {journal}
  {\bibinfo  {journal} {Nature}\ }\textbf {\bibinfo {volume} {583}},\ \bibinfo
  {pages} {369} (\bibinfo {year} {2020})}\BibitemShut {NoStop}%
\bibitem [{\citenamefont {Xing}\ \emph
  {et~al.}(2024{\natexlab{a}})\citenamefont {Xing}, \citenamefont {Wei},\ and\
  \citenamefont {Liao}}]{xing2024a}%
  \BibitemOpen
  \bibfield  {author} {\bibinfo {author} {\bibfnamefont {F.}~\bibnamefont
  {Xing}}, \bibinfo {author} {\bibfnamefont {Y.}~\bibnamefont {Wei}},\ and\
  \bibinfo {author} {\bibfnamefont {Z.}~\bibnamefont {Liao}},\ }\bibfield
  {title} {\bibinfo {title} {Quantum search in many-body interacting systems
  with long-range interactions},\ }\href
  {https://doi.org/10.1103/PhysRevA.109.052435} {\bibfield  {journal} {\bibinfo
   {journal} {Phys. Rev. A}\ }\textbf {\bibinfo {volume} {109}},\ \bibinfo
  {pages} {052435} (\bibinfo {year} {2024}{\natexlab{a}})}\BibitemShut
  {NoStop}%
\bibitem [{\citenamefont {Xing}\ \emph
  {et~al.}(2024{\natexlab{b}})\citenamefont {Xing}, \citenamefont {Liao},\ and\
  \citenamefont {Wang}}]{xing2024}%
  \BibitemOpen
  \bibfield  {author} {\bibinfo {author} {\bibfnamefont {F.}~\bibnamefont
  {Xing}}, \bibinfo {author} {\bibfnamefont {Z.}~\bibnamefont {Liao}},\ and\
  \bibinfo {author} {\bibfnamefont {X.-h.}\ \bibnamefont {Wang}},\ }\bibfield
  {title} {\bibinfo {title} {Deterministic generation of arbitrary n-photon
  states in a waveguide-{QED} system},\ }\href
  {https://doi.org/10.1103/PhysRevA.109.013718} {\bibfield  {journal} {\bibinfo
   {journal} {Phys. Rev. A}\ }\textbf {\bibinfo {volume} {109}},\ \bibinfo
  {pages} {013718} (\bibinfo {year} {2024}{\natexlab{b}})}\BibitemShut
  {NoStop}%
\bibitem [{\citenamefont {Lu}\ \emph {et~al.}(2025)\citenamefont {Lu},
  \citenamefont {Liao},\ and\ \citenamefont {Wang}}]{lu2025}%
  \BibitemOpen
  \bibfield  {author} {\bibinfo {author} {\bibfnamefont {Y.}~\bibnamefont
  {Lu}}, \bibinfo {author} {\bibfnamefont {Z.}~\bibnamefont {Liao}},\ and\
  \bibinfo {author} {\bibfnamefont {X.-H.}\ \bibnamefont {Wang}},\ }\bibfield
  {title} {\bibinfo {title} {Atomic-scale on-demand photon polarization
  manipulation with high efficiency for integrated photonic chips},\ }\href
  {https://doi.org/10.1103/PhysRevLett.134.083601} {\bibfield  {journal}
  {\bibinfo  {journal} {Phys. Rev. Lett.}\ }\textbf {\bibinfo {volume} {134}},\
  \bibinfo {pages} {083601} (\bibinfo {year} {2025})}\BibitemShut {NoStop}%
\bibitem [{\citenamefont {{Gonz{\'a}lez-Tudela}}\ \emph
  {et~al.}(2015)\citenamefont {{Gonz{\'a}lez-Tudela}}, \citenamefont
  {Paulisch}, \citenamefont {Chang}, \citenamefont {Kimble},\ and\
  \citenamefont {Cirac}}]{gonzalez-tudela2015}%
  \BibitemOpen
  \bibfield  {author} {\bibinfo {author} {\bibfnamefont {A.}~\bibnamefont
  {{Gonz{\'a}lez-Tudela}}}, \bibinfo {author} {\bibfnamefont {V.}~\bibnamefont
  {Paulisch}}, \bibinfo {author} {\bibfnamefont {D.~E.}\ \bibnamefont {Chang}},
  \bibinfo {author} {\bibfnamefont {H.~J.}\ \bibnamefont {Kimble}},\ and\
  \bibinfo {author} {\bibfnamefont {J.~I.}\ \bibnamefont {Cirac}},\ }\bibfield
  {title} {\bibinfo {title} {Deterministic generation of arbitrary photonic
  states assisted by dissipation},\ }\href
  {https://doi.org/10.1103/PhysRevLett.115.163603} {\bibfield  {journal}
  {\bibinfo  {journal} {Phys. Rev. Lett.}\ }\textbf {\bibinfo {volume} {115}},\
  \bibinfo {pages} {163603} (\bibinfo {year} {2015})}\BibitemShut {NoStop}%
\bibitem [{\citenamefont {{Gonz{\'a}lez-Tudela}}\ \emph
  {et~al.}(2017)\citenamefont {{Gonz{\'a}lez-Tudela}}, \citenamefont
  {Paulisch}, \citenamefont {Kimble},\ and\ \citenamefont
  {Cirac}}]{gonzalez-tudela2017}%
  \BibitemOpen
  \bibfield  {author} {\bibinfo {author} {\bibfnamefont {A.}~\bibnamefont
  {{Gonz{\'a}lez-Tudela}}}, \bibinfo {author} {\bibfnamefont {V.}~\bibnamefont
  {Paulisch}}, \bibinfo {author} {\bibfnamefont {H.~J.}\ \bibnamefont
  {Kimble}},\ and\ \bibinfo {author} {\bibfnamefont {J.~I.}\ \bibnamefont
  {Cirac}},\ }\bibfield  {title} {\bibinfo {title} {Efficient multiphoton
  generation in waveguide quantum electrodynamics},\ }\href
  {https://doi.org/10.1103/PhysRevLett.118.213601} {\bibfield  {journal}
  {\bibinfo  {journal} {Phys. Rev. Lett.}\ }\textbf {\bibinfo {volume} {118}},\
  \bibinfo {pages} {213601} (\bibinfo {year} {2017})}\BibitemShut {NoStop}%
\bibitem [{\citenamefont {Ke}\ \emph {et~al.}(2019)\citenamefont {Ke},
  \citenamefont {Poshakinskiy}, \citenamefont {Lee}, \citenamefont {Kivshar},\
  and\ \citenamefont {Poddubny}}]{ke2019}%
  \BibitemOpen
  \bibfield  {author} {\bibinfo {author} {\bibfnamefont {Y.}~\bibnamefont
  {Ke}}, \bibinfo {author} {\bibfnamefont {A.~V.}\ \bibnamefont
  {Poshakinskiy}}, \bibinfo {author} {\bibfnamefont {C.}~\bibnamefont {Lee}},
  \bibinfo {author} {\bibfnamefont {Y.~S.}\ \bibnamefont {Kivshar}},\ and\
  \bibinfo {author} {\bibfnamefont {A.~N.}\ \bibnamefont {Poddubny}},\
  }\bibfield  {title} {\bibinfo {title} {Inelastic scattering of photon pairs
  in qubit arrays with subradiant states},\ }\href
  {https://doi.org/10.1103/PhysRevLett.123.253601} {\bibfield  {journal}
  {\bibinfo  {journal} {Phys. Rev. Lett.}\ }\textbf {\bibinfo {volume} {123}},\
  \bibinfo {pages} {253601} (\bibinfo {year} {2019})}\BibitemShut {NoStop}%
\bibitem [{\citenamefont {Wang}\ and\ \citenamefont {Jia}(2024)}]{wang2024b}%
  \BibitemOpen
  \bibfield  {author} {\bibinfo {author} {\bibfnamefont {M.~S.}\ \bibnamefont
  {Wang}}\ and\ \bibinfo {author} {\bibfnamefont {W.~Z.}\ \bibnamefont {Jia}},\
  }\bibfield  {title} {\bibinfo {title} {Engineering photonic band gaps with a
  waveguide-{QED} structure containing an atom-polymer array},\ }\href
  {https://doi.org/10.1103/PhysRevA.110.053716} {\bibfield  {journal} {\bibinfo
   {journal} {Phys. Rev. A}\ }\textbf {\bibinfo {volume} {110}},\ \bibinfo
  {pages} {053716} (\bibinfo {year} {2024})}\BibitemShut {NoStop}%
\bibitem [{\citenamefont {Wang}\ \emph {et~al.}(2025)\citenamefont {Wang},
  \citenamefont {He}, \citenamefont {Liao},\ and\ \citenamefont
  {Zubairy}}]{wang2025}%
  \BibitemOpen
  \bibfield  {author} {\bibinfo {author} {\bibfnamefont {X.}~\bibnamefont
  {Wang}}, \bibinfo {author} {\bibfnamefont {J.}~\bibnamefont {He}}, \bibinfo
  {author} {\bibfnamefont {Z.}~\bibnamefont {Liao}},\ and\ \bibinfo {author}
  {\bibfnamefont {M.~S.}\ \bibnamefont {Zubairy}},\ }\bibfield  {title}
  {\bibinfo {title} {Tunable ultrahigh broadband reflection via collective
  atom-atom interaction in a waveguide-{QED} system},\ }\href
  {https://doi.org/10.1103/PhysRevA.111.013706} {\bibfield  {journal} {\bibinfo
   {journal} {Phys. Rev. A}\ }\textbf {\bibinfo {volume} {111}},\ \bibinfo
  {pages} {013706} (\bibinfo {year} {2025})}\BibitemShut {NoStop}%
\bibitem [{\citenamefont {Liao}\ \emph {et~al.}(2017)\citenamefont {Liao},
  \citenamefont {{Al-Amri}},\ and\ \citenamefont {Zubairy}}]{liao2017}%
  \BibitemOpen
  \bibfield  {author} {\bibinfo {author} {\bibfnamefont {Z.}~\bibnamefont
  {Liao}}, \bibinfo {author} {\bibfnamefont {M.}~\bibnamefont {{Al-Amri}}},\
  and\ \bibinfo {author} {\bibfnamefont {M.~S.}\ \bibnamefont {Zubairy}},\
  }\bibfield  {title} {\bibinfo {title} {Measurement of deep-subwavelength
  emitter separation in a waveguide-{QED} system},\ }\href
  {https://doi.org/10.1364/OE.25.031997} {\bibfield  {journal} {\bibinfo
  {journal} {Opt. Express}\ }\textbf {\bibinfo {volume} {25}},\ \bibinfo
  {pages} {31997} (\bibinfo {year} {2017})}\BibitemShut {NoStop}%
\bibitem [{\citenamefont {Zhou}\ \emph
  {et~al.}(2025{\natexlab{a}})\citenamefont {Zhou}, \citenamefont {Chang},\
  and\ \citenamefont {Yang}}]{zhou2025}%
  \BibitemOpen
  \bibfield  {author} {\bibinfo {author} {\bibfnamefont {T.-Q.}\ \bibnamefont
  {Zhou}}, \bibinfo {author} {\bibfnamefont {Y.}~\bibnamefont {Chang}},\ and\
  \bibinfo {author} {\bibfnamefont {L.-P.}\ \bibnamefont {Yang}},\ }\bibfield
  {title} {\bibinfo {title} {Dark-state-induced transparency and the
  ultranarrow spectrum},\ }\href {https://doi.org/10.1103/PhysRevA.111.033708}
  {\bibfield  {journal} {\bibinfo  {journal} {Phys. Rev. A}\ }\textbf {\bibinfo
  {volume} {111}},\ \bibinfo {pages} {033708} (\bibinfo {year}
  {2025}{\natexlab{a}})}\BibitemShut {NoStop}%
\bibitem [{\citenamefont {{Asenjo-Garcia}}\ \emph {et~al.}(2017)\citenamefont
  {{Asenjo-Garcia}}, \citenamefont {Hood}, \citenamefont {Chang},\ and\
  \citenamefont {Kimble}}]{asenjo-garcia2017}%
  \BibitemOpen
  \bibfield  {author} {\bibinfo {author} {\bibfnamefont {A.}~\bibnamefont
  {{Asenjo-Garcia}}}, \bibinfo {author} {\bibfnamefont {J.~D.}\ \bibnamefont
  {Hood}}, \bibinfo {author} {\bibfnamefont {D.~E.}\ \bibnamefont {Chang}},\
  and\ \bibinfo {author} {\bibfnamefont {H.~J.}\ \bibnamefont {Kimble}},\
  }\bibfield  {title} {\bibinfo {title} {Atom-light interactions in
  quasi-one-dimensional nanostructures: A green's-function perspective},\
  }\href {https://doi.org/10.1103/PhysRevA.95.033818} {\bibfield  {journal}
  {\bibinfo  {journal} {Phys. Rev. A}\ }\textbf {\bibinfo {volume} {95}},\
  \bibinfo {pages} {033818} (\bibinfo {year} {2017})}\BibitemShut {NoStop}%
\bibitem [{\citenamefont {Zhang}\ \emph {et~al.}(2020)\citenamefont {Zhang},
  \citenamefont {Yu},\ and\ \citenamefont {M{\o}lmer}}]{zhang2020b}%
  \BibitemOpen
  \bibfield  {author} {\bibinfo {author} {\bibfnamefont {Y.-X.}\ \bibnamefont
  {Zhang}}, \bibinfo {author} {\bibfnamefont {C.}~\bibnamefont {Yu}},\ and\
  \bibinfo {author} {\bibfnamefont {K.}~\bibnamefont {M{\o}lmer}},\ }\bibfield
  {title} {\bibinfo {title} {Subradiant bound dimer excited states of emitter
  chains coupled to a one dimensional waveguide},\ }\href
  {https://doi.org/10.1103/PhysRevResearch.2.013173} {\bibfield  {journal}
  {\bibinfo  {journal} {Phys. Rev. Res.}\ }\textbf {\bibinfo {volume} {2}},\
  \bibinfo {pages} {013173} (\bibinfo {year} {2020})}\BibitemShut {NoStop}%
\bibitem [{\citenamefont {Kornovan}\ \emph {et~al.}(2019)\citenamefont
  {Kornovan}, \citenamefont {Corzo}, \citenamefont {Laurat},\ and\
  \citenamefont {Sheremet}}]{kornovan2019}%
  \BibitemOpen
  \bibfield  {author} {\bibinfo {author} {\bibfnamefont {D.~F.}\ \bibnamefont
  {Kornovan}}, \bibinfo {author} {\bibfnamefont {N.~V.}\ \bibnamefont {Corzo}},
  \bibinfo {author} {\bibfnamefont {J.}~\bibnamefont {Laurat}},\ and\ \bibinfo
  {author} {\bibfnamefont {A.~S.}\ \bibnamefont {Sheremet}},\ }\bibfield
  {title} {\bibinfo {title} {Extremely subradiant states in a periodic
  one-dimensional atomic array},\ }\href
  {https://doi.org/10.1103/PhysRevA.100.063832} {\bibfield  {journal} {\bibinfo
   {journal} {Phys. Rev. A}\ }\textbf {\bibinfo {volume} {100}},\ \bibinfo
  {pages} {063832} (\bibinfo {year} {2019})}\BibitemShut {NoStop}%
\bibitem [{\citenamefont {Liao}\ \emph
  {et~al.}(2016{\natexlab{b}})\citenamefont {Liao}, \citenamefont {Nha},\ and\
  \citenamefont {Zubairy}}]{liao2016}%
  \BibitemOpen
  \bibfield  {author} {\bibinfo {author} {\bibfnamefont {Z.}~\bibnamefont
  {Liao}}, \bibinfo {author} {\bibfnamefont {H.}~\bibnamefont {Nha}},\ and\
  \bibinfo {author} {\bibfnamefont {M.~S.}\ \bibnamefont {Zubairy}},\
  }\bibfield  {title} {\bibinfo {title} {Dynamical theory of single-photon
  transport in a one-dimensional waveguide coupled to identical and
  nonidentical emitters},\ }\href {https://doi.org/10.1103/PhysRevA.94.053842}
  {\bibfield  {journal} {\bibinfo  {journal} {Phys. Rev. A}\ }\textbf {\bibinfo
  {volume} {94}},\ \bibinfo {pages} {053842} (\bibinfo {year}
  {2016}{\natexlab{b}})}\BibitemShut {NoStop}%
\bibitem [{\citenamefont {Andreoli}\ \emph {et~al.}(2023)\citenamefont
  {Andreoli}, \citenamefont {Windt}, \citenamefont {Grava}, \citenamefont
  {Andolina}, \citenamefont {Gullans}, \citenamefont {High},\ and\
  \citenamefont {Chang}}]{andreoli2023}%
  \BibitemOpen
  \bibfield  {author} {\bibinfo {author} {\bibfnamefont {F.}~\bibnamefont
  {Andreoli}}, \bibinfo {author} {\bibfnamefont {B.}~\bibnamefont {Windt}},
  \bibinfo {author} {\bibfnamefont {S.}~\bibnamefont {Grava}}, \bibinfo
  {author} {\bibfnamefont {G.~M.}\ \bibnamefont {Andolina}}, \bibinfo {author}
  {\bibfnamefont {M.~J.}\ \bibnamefont {Gullans}}, \bibinfo {author}
  {\bibfnamefont {A.~A.}\ \bibnamefont {High}},\ and\ \bibinfo {author}
  {\bibfnamefont {D.~E.}\ \bibnamefont {Chang}},\ }\href
  {https://arxiv.org/abs/2303.10998} {\bibinfo {title} {The maximum refractive
  index of an atomic crystal : from quantum optics to quantum chemistry}}
  (\bibinfo {year} {2023}),\ \bibinfo {note} {arxiv:2303.10998v1},\ \Eprint
  {https://arxiv.org/abs/2303.10998} {arXiv:2303.10998 [quant-ph]} \BibitemShut
  {NoStop}%
\bibitem [{\citenamefont {{A. Asenjo-Garcia}}\ \emph
  {et~al.}(2017)\citenamefont {{A. Asenjo-Garcia}}, \citenamefont {{M.
  Moreno-Cardoner}}, \citenamefont {{A. Albrecht}}, \citenamefont {{H. J.
  Kimble}},\ and\ \citenamefont {{D. E. Chang}}}]{a.asenjo-garcia2017}%
  \BibitemOpen
  \bibfield  {author} {\bibinfo {author} {\bibnamefont {{A. Asenjo-Garcia}}},
  \bibinfo {author} {\bibnamefont {{M. Moreno-Cardoner}}}, \bibinfo {author}
  {\bibnamefont {{A. Albrecht}}}, \bibinfo {author} {\bibnamefont {{H. J.
  Kimble}}},\ and\ \bibinfo {author} {\bibnamefont {{D. E. Chang}}},\
  }\bibfield  {title} {\bibinfo {title} {Exponential improvement in photon
  storage fidelities using subradiance and ``selective radiance'' in atomic
  arrays},\ }\href {https://doi.org/10.1103/PhysRevX.7.031024} {\bibfield
  {journal} {\bibinfo  {journal} {Phys. Rev. X}\ }\textbf {\bibinfo {volume}
  {7}},\ \bibinfo {pages} {031024} (\bibinfo {year} {2017})}\BibitemShut
  {NoStop}%
\bibitem [{sup()}]{supplementary}%
  \BibitemOpen
  \href@noop {} {}\bibinfo {note} {See Supplemental Material at [URL will be
  inserted by publisher] for detailed derivations of the subradiant decay-rate
  formulas, additional spectral and eigenvalue results, guided-mode phase
  mismatch, and asymmetric and chiral guided couplings.}\BibitemShut {Stop}%
\bibitem [{\citenamefont {Polino}\ \emph {et~al.}(2020)\citenamefont {Polino},
  \citenamefont {Valeri}, \citenamefont {Spagnolo},\ and\ \citenamefont
  {Sciarrino}}]{Polino2020}%
  \BibitemOpen
  \bibfield  {author} {\bibinfo {author} {\bibfnamefont {E.}~\bibnamefont
  {Polino}}, \bibinfo {author} {\bibfnamefont {M.}~\bibnamefont {Valeri}},
  \bibinfo {author} {\bibfnamefont {N.}~\bibnamefont {Spagnolo}},\ and\
  \bibinfo {author} {\bibfnamefont {F.}~\bibnamefont {Sciarrino}},\ }\bibfield
  {title} {\bibinfo {title} {Photonic quantum metrology},\ }\href
  {https://doi.org/10.1116/5.0007577} {\bibfield  {journal} {\bibinfo
  {journal} {AVS Quantum Sci.}\ }\textbf {\bibinfo {volume} {2}},\ \bibinfo
  {pages} {024703} (\bibinfo {year} {2020})}\BibitemShut {NoStop}%
\bibitem [{\citenamefont {Corzo}\ \emph {et~al.}(2019)\citenamefont {Corzo},
  \citenamefont {Raskop}, \citenamefont {Chandra}, \citenamefont {Sheremet},
  \citenamefont {Gouraud},\ and\ \citenamefont {Laurat}}]{corzo2019}%
  \BibitemOpen
  \bibfield  {author} {\bibinfo {author} {\bibfnamefont {N.~V.}\ \bibnamefont
  {Corzo}}, \bibinfo {author} {\bibfnamefont {J.}~\bibnamefont {Raskop}},
  \bibinfo {author} {\bibfnamefont {A.}~\bibnamefont {Chandra}}, \bibinfo
  {author} {\bibfnamefont {A.~S.}\ \bibnamefont {Sheremet}}, \bibinfo {author}
  {\bibfnamefont {B.}~\bibnamefont {Gouraud}},\ and\ \bibinfo {author}
  {\bibfnamefont {J.}~\bibnamefont {Laurat}},\ }\bibfield  {title} {\bibinfo
  {title} {Waveguide-coupled single collective excitation of atomic arrays},\
  }\href {https://doi.org/10.1038/s41586-019-0902-3} {\bibfield  {journal}
  {\bibinfo  {journal} {Nature}\ }\textbf {\bibinfo {volume} {566}},\ \bibinfo
  {pages} {359} (\bibinfo {year} {2019})}\BibitemShut {NoStop}%
\bibitem [{\citenamefont {Sunami}\ \emph {et~al.}(2025)\citenamefont {Sunami},
  \citenamefont {Tamiya}, \citenamefont {Inoue}, \citenamefont {Yamasaki},\
  and\ \citenamefont {Goban}}]{sunami2025}%
  \BibitemOpen
  \bibfield  {author} {\bibinfo {author} {\bibfnamefont {S.}~\bibnamefont
  {Sunami}}, \bibinfo {author} {\bibfnamefont {S.}~\bibnamefont {Tamiya}},
  \bibinfo {author} {\bibfnamefont {R.}~\bibnamefont {Inoue}}, \bibinfo
  {author} {\bibfnamefont {H.}~\bibnamefont {Yamasaki}},\ and\ \bibinfo
  {author} {\bibfnamefont {A.}~\bibnamefont {Goban}},\ }\bibfield  {title}
  {\bibinfo {title} {Scalable networking of neutral-atom qubits:
  Nanofiber-based approach for multiprocessor fault-tolerant quantum
  computers},\ }\href {https://doi.org/10.1103/PRXQuantum.6.010101} {\bibfield
  {journal} {\bibinfo  {journal} {PRX Quantum}\ }\textbf {\bibinfo {volume}
  {6}},\ \bibinfo {pages} {010101} (\bibinfo {year} {2025})}\BibitemShut
  {NoStop}%
\bibitem [{\citenamefont {Zhou}\ \emph
  {et~al.}(2025{\natexlab{b}})\citenamefont {Zhou}, \citenamefont {Suresh},
  \citenamefont {Robicheaux},\ and\ \citenamefont {Hung}}]{zhou2025selective}%
  \BibitemOpen
  \bibfield  {author} {\bibinfo {author} {\bibfnamefont {X.}~\bibnamefont
  {Zhou}}, \bibinfo {author} {\bibfnamefont {D.~A.}\ \bibnamefont {Suresh}},
  \bibinfo {author} {\bibfnamefont {F.}~\bibnamefont {Robicheaux}},\ and\
  \bibinfo {author} {\bibfnamefont {C.-L.}\ \bibnamefont {Hung}},\ }\bibfield
  {title} {\bibinfo {title} {Selective collective emission from a dense atomic
  ensemble coupled to a nanophotonic resonator},\ }\href
  {https://doi.org/10.1103/PhysRevLett.135.113601} {\bibfield  {journal}
  {\bibinfo  {journal} {Physical Review Letters}\ }\textbf {\bibinfo {volume}
  {135}},\ \bibinfo {pages} {113601} (\bibinfo {year}
  {2025}{\natexlab{b}})}\BibitemShut {NoStop}%
\bibitem [{\citenamefont {Takahata}\ \emph {et~al.}(2026)\citenamefont
  {Takahata}, \citenamefont {Keloth}, \citenamefont {Yamamoto}, \citenamefont
  {Harada}, \citenamefont {Miki},\ and\ \citenamefont
  {Aoki}}]{takahata2026fiber}%
  \BibitemOpen
  \bibfield  {author} {\bibinfo {author} {\bibfnamefont {M.}~\bibnamefont
  {Takahata}}, \bibinfo {author} {\bibfnamefont {J.}~\bibnamefont {Keloth}},
  \bibinfo {author} {\bibfnamefont {T.}~\bibnamefont {Yamamoto}}, \bibinfo
  {author} {\bibfnamefont {K.-i.}\ \bibnamefont {Harada}}, \bibinfo {author}
  {\bibfnamefont {S.}~\bibnamefont {Miki}},\ and\ \bibinfo {author}
  {\bibfnamefont {T.}~\bibnamefont {Aoki}},\ }\bibfield  {title} {\bibinfo
  {title} {Fiber-optic quantum interface with an array of more than 100
  individually addressable atoms on an optical nanofiber},\ }\href@noop {}
  {\bibfield  {journal} {\bibinfo  {journal} {arXiv preprint arXiv:2603.21812}\
  } (\bibinfo {year} {2026})}\BibitemShut {NoStop}%
\bibitem [{\citenamefont {Pache}\ \emph {et~al.}(2025)\citenamefont {Pache},
  \citenamefont {Cordier}, \citenamefont {Letellier}, \citenamefont {Schemmer},
  \citenamefont {Schneeweiss}, \citenamefont {Volz},\ and\ \citenamefont
  {Rauschenbeutel}}]{pache2025magic}%
  \BibitemOpen
  \bibfield  {author} {\bibinfo {author} {\bibfnamefont {L.}~\bibnamefont
  {Pache}}, \bibinfo {author} {\bibfnamefont {M.}~\bibnamefont {Cordier}},
  \bibinfo {author} {\bibfnamefont {H.}~\bibnamefont {Letellier}}, \bibinfo
  {author} {\bibfnamefont {M.}~\bibnamefont {Schemmer}}, \bibinfo {author}
  {\bibfnamefont {P.}~\bibnamefont {Schneeweiss}}, \bibinfo {author}
  {\bibfnamefont {J.}~\bibnamefont {Volz}},\ and\ \bibinfo {author}
  {\bibfnamefont {A.}~\bibnamefont {Rauschenbeutel}},\ }\bibfield  {title}
  {\bibinfo {title} {Magic-wavelength nanofiber-based two-color dipole trap
  with sub-$\lambda/2$ spacing},\ }\href
  {https://doi.org/10.1103/PhysRevA.112.L011701} {\bibfield  {journal}
  {\bibinfo  {journal} {Physical Review A}\ }\textbf {\bibinfo {volume}
  {112}},\ \bibinfo {pages} {L011701} (\bibinfo {year} {2025})}\BibitemShut
  {NoStop}%
\bibitem [{\citenamefont {Holzinger}\ \emph {et~al.}(2022)\citenamefont
  {Holzinger}, \citenamefont {Oh}, \citenamefont {Reitz}, \citenamefont
  {Ritsch},\ and\ \citenamefont {Genes}}]{holzinger2022cooperative}%
  \BibitemOpen
  \bibfield  {author} {\bibinfo {author} {\bibfnamefont {R.}~\bibnamefont
  {Holzinger}}, \bibinfo {author} {\bibfnamefont {S.~A.}\ \bibnamefont {Oh}},
  \bibinfo {author} {\bibfnamefont {M.}~\bibnamefont {Reitz}}, \bibinfo
  {author} {\bibfnamefont {H.}~\bibnamefont {Ritsch}},\ and\ \bibinfo {author}
  {\bibfnamefont {C.}~\bibnamefont {Genes}},\ }\bibfield  {title} {\bibinfo
  {title} {Cooperative subwavelength molecular quantum emitter arrays},\ }\href
  {https://doi.org/10.1103/PhysRevResearch.4.033116} {\bibfield  {journal}
  {\bibinfo  {journal} {Physical Review Research}\ }\textbf {\bibinfo {volume}
  {4}},\ \bibinfo {pages} {033116} (\bibinfo {year} {2022})}\BibitemShut
  {NoStop}%
\bibitem [{\citenamefont {Seubert}\ \emph {et~al.}(2025)\citenamefont
  {Seubert}, \citenamefont {Hartung}, \citenamefont {Welte}, \citenamefont
  {Rempe},\ and\ \citenamefont {Distante}}]{Seubert2025}%
  \BibitemOpen
  \bibfield  {author} {\bibinfo {author} {\bibfnamefont {M.}~\bibnamefont
  {Seubert}}, \bibinfo {author} {\bibfnamefont {L.}~\bibnamefont {Hartung}},
  \bibinfo {author} {\bibfnamefont {S.}~\bibnamefont {Welte}}, \bibinfo
  {author} {\bibfnamefont {G.}~\bibnamefont {Rempe}},\ and\ \bibinfo {author}
  {\bibfnamefont {E.}~\bibnamefont {Distante}},\ }\bibfield  {title} {\bibinfo
  {title} {Tweezer-assisted subwavelength positioning of atomic arrays in an
  optical cavity},\ }\href {https://doi.org/10.1103/PRXQuantum.6.010322}
  {\bibfield  {journal} {\bibinfo  {journal} {PRX Quantum}\ }\textbf {\bibinfo
  {volume} {6}},\ \bibinfo {pages} {010322} (\bibinfo {year}
  {2025})}\BibitemShut {NoStop}%
\bibitem [{\citenamefont {Srakaew}\ \emph {et~al.}(2023)\citenamefont
  {Srakaew}, \citenamefont {Weckesser}, \citenamefont {Hollerith},
  \citenamefont {Wei}, \citenamefont {Adler}, \citenamefont {Bloch},\ and\
  \citenamefont {Zeiher}}]{Srakaew2023}%
  \BibitemOpen
  \bibfield  {author} {\bibinfo {author} {\bibfnamefont {K.}~\bibnamefont
  {Srakaew}}, \bibinfo {author} {\bibfnamefont {P.}~\bibnamefont {Weckesser}},
  \bibinfo {author} {\bibfnamefont {S.}~\bibnamefont {Hollerith}}, \bibinfo
  {author} {\bibfnamefont {D.}~\bibnamefont {Wei}}, \bibinfo {author}
  {\bibfnamefont {D.}~\bibnamefont {Adler}}, \bibinfo {author} {\bibfnamefont
  {I.}~\bibnamefont {Bloch}},\ and\ \bibinfo {author} {\bibfnamefont
  {J.}~\bibnamefont {Zeiher}},\ }\bibfield  {title} {\bibinfo {title} {A
  subwavelength atomic array switched by a single rydberg atom},\ }\href
  {https://doi.org/10.1038/s41567-023-01959-y} {\bibfield  {journal} {\bibinfo
  {journal} {Nature Physics}\ }\textbf {\bibinfo {volume} {19}},\ \bibinfo
  {pages} {714} (\bibinfo {year} {2023})}\BibitemShut {NoStop}%
\bibitem [{\citenamefont {Sipahigil}\ \emph {et~al.}(2016)\citenamefont
  {Sipahigil}, \citenamefont {Evans}, \citenamefont {Sukachev}, \citenamefont
  {Burek}, \citenamefont {Borregaard}, \citenamefont {Bhaskar}, \citenamefont
  {Nguyen}, \citenamefont {Pacheco}, \citenamefont {Atikian}, \citenamefont
  {Meuwly}, \citenamefont {Camacho}, \citenamefont {Jelezko}, \citenamefont
  {Bielejec}, \citenamefont {Park}, \citenamefont {Lon{\v c}ar},\ and\
  \citenamefont {Lukin}}]{sipahigil2016}%
  \BibitemOpen
  \bibfield  {author} {\bibinfo {author} {\bibfnamefont {A.}~\bibnamefont
  {Sipahigil}}, \bibinfo {author} {\bibfnamefont {R.~E.}\ \bibnamefont
  {Evans}}, \bibinfo {author} {\bibfnamefont {D.~D.}\ \bibnamefont {Sukachev}},
  \bibinfo {author} {\bibfnamefont {M.~J.}\ \bibnamefont {Burek}}, \bibinfo
  {author} {\bibfnamefont {J.}~\bibnamefont {Borregaard}}, \bibinfo {author}
  {\bibfnamefont {M.~K.}\ \bibnamefont {Bhaskar}}, \bibinfo {author}
  {\bibfnamefont {C.~T.}\ \bibnamefont {Nguyen}}, \bibinfo {author}
  {\bibfnamefont {J.~L.}\ \bibnamefont {Pacheco}}, \bibinfo {author}
  {\bibfnamefont {H.~A.}\ \bibnamefont {Atikian}}, \bibinfo {author}
  {\bibfnamefont {C.}~\bibnamefont {Meuwly}}, \bibinfo {author} {\bibfnamefont
  {R.~M.}\ \bibnamefont {Camacho}}, \bibinfo {author} {\bibfnamefont
  {F.}~\bibnamefont {Jelezko}}, \bibinfo {author} {\bibfnamefont
  {E.}~\bibnamefont {Bielejec}}, \bibinfo {author} {\bibfnamefont
  {H.}~\bibnamefont {Park}}, \bibinfo {author} {\bibfnamefont {M.}~\bibnamefont
  {Lon{\v c}ar}},\ and\ \bibinfo {author} {\bibfnamefont {M.~D.}\ \bibnamefont
  {Lukin}},\ }\bibfield  {title} {\bibinfo {title} {An integrated diamond
  nanophotonics platform for quantum-optical networks},\ }\href
  {https://doi.org/10.1126/science.aah6875} {\bibfield  {journal} {\bibinfo
  {journal} {Science}\ }\textbf {\bibinfo {volume} {354}},\ \bibinfo {pages}
  {847} (\bibinfo {year} {2016})}\BibitemShut {NoStop}%
\bibitem [{\citenamefont {Ngan}\ \emph {et~al.}(2023)\citenamefont {Ngan},
  \citenamefont {Zhan}, \citenamefont {Dory}, \citenamefont {Vu{\v
  c}kovi{\'c}},\ and\ \citenamefont {Sun}}]{ngan2023}%
  \BibitemOpen
  \bibfield  {author} {\bibinfo {author} {\bibfnamefont {K.}~\bibnamefont
  {Ngan}}, \bibinfo {author} {\bibfnamefont {Y.}~\bibnamefont {Zhan}}, \bibinfo
  {author} {\bibfnamefont {C.}~\bibnamefont {Dory}}, \bibinfo {author}
  {\bibfnamefont {J.}~\bibnamefont {Vu{\v c}kovi{\'c}}},\ and\ \bibinfo
  {author} {\bibfnamefont {S.}~\bibnamefont {Sun}},\ }\bibfield  {title}
  {\bibinfo {title} {Quantum photonic circuits integrated with color centers in
  designer nanodiamonds},\ }\href
  {https://doi.org/10.1021/acs.nanolett.3c02645} {\bibfield  {journal}
  {\bibinfo  {journal} {Nano Lett.}\ }\textbf {\bibinfo {volume} {23}},\
  \bibinfo {pages} {9360} (\bibinfo {year} {2023})}\BibitemShut {NoStop}%
\bibitem [{\citenamefont {Tzeng}\ \emph {et~al.}(2024)\citenamefont {Tzeng},
  \citenamefont {Ke}, \citenamefont {Jia}, \citenamefont {Liu}, \citenamefont
  {Park}, \citenamefont {Han}, \citenamefont {Frost}, \citenamefont {Cai},
  \citenamefont {Mao}, \citenamefont {Ewing}, \citenamefont {Cui},
  \citenamefont {Devereaux}, \citenamefont {Lin},\ and\ \citenamefont
  {Chu}}]{tzeng2024}%
  \BibitemOpen
  \bibfield  {author} {\bibinfo {author} {\bibfnamefont {Y.-K.}\ \bibnamefont
  {Tzeng}}, \bibinfo {author} {\bibfnamefont {F.}~\bibnamefont {Ke}}, \bibinfo
  {author} {\bibfnamefont {C.}~\bibnamefont {Jia}}, \bibinfo {author}
  {\bibfnamefont {Y.}~\bibnamefont {Liu}}, \bibinfo {author} {\bibfnamefont
  {S.}~\bibnamefont {Park}}, \bibinfo {author} {\bibfnamefont {M.}~\bibnamefont
  {Han}}, \bibinfo {author} {\bibfnamefont {M.}~\bibnamefont {Frost}}, \bibinfo
  {author} {\bibfnamefont {X.}~\bibnamefont {Cai}}, \bibinfo {author}
  {\bibfnamefont {W.~L.}\ \bibnamefont {Mao}}, \bibinfo {author} {\bibfnamefont
  {R.~C.}\ \bibnamefont {Ewing}}, \bibinfo {author} {\bibfnamefont
  {Y.}~\bibnamefont {Cui}}, \bibinfo {author} {\bibfnamefont {T.~P.}\
  \bibnamefont {Devereaux}}, \bibinfo {author} {\bibfnamefont {Y.}~\bibnamefont
  {Lin}},\ and\ \bibinfo {author} {\bibfnamefont {S.}~\bibnamefont {Chu}},\
  }\bibfield  {title} {\bibinfo {title} {Improving the creation of {SiV}
  centers in diamond via sub-{$\mu$}s pulsed annealing treatment},\ }\href
  {https://doi.org/10.1038/s41467-024-51523-2} {\bibfield  {journal} {\bibinfo
  {journal} {Nat. Commun.}\ }\textbf {\bibinfo {volume} {15}},\ \bibinfo
  {pages} {7251} (\bibinfo {year} {2024})}\BibitemShut {NoStop}%
\bibitem [{\citenamefont {Day}\ \emph {et~al.}(2022)\citenamefont {Day},
  \citenamefont {Bates}, \citenamefont {Smallwood}, \citenamefont {Owen},
  \citenamefont {Schr{\"o}der}, \citenamefont {Bielejec}, \citenamefont
  {Ulbricht},\ and\ \citenamefont {Cundiff}}]{day2022coherent}%
  \BibitemOpen
  \bibfield  {author} {\bibinfo {author} {\bibfnamefont {M.~W.}\ \bibnamefont
  {Day}}, \bibinfo {author} {\bibfnamefont {K.~M.}\ \bibnamefont {Bates}},
  \bibinfo {author} {\bibfnamefont {C.~L.}\ \bibnamefont {Smallwood}}, \bibinfo
  {author} {\bibfnamefont {R.~C.}\ \bibnamefont {Owen}}, \bibinfo {author}
  {\bibfnamefont {T.}~\bibnamefont {Schr{\"o}der}}, \bibinfo {author}
  {\bibfnamefont {E.}~\bibnamefont {Bielejec}}, \bibinfo {author}
  {\bibfnamefont {R.}~\bibnamefont {Ulbricht}},\ and\ \bibinfo {author}
  {\bibfnamefont {S.~T.}\ \bibnamefont {Cundiff}},\ }\bibfield  {title}
  {\bibinfo {title} {Coherent interactions between silicon-vacancy centers in
  diamond},\ }\href {https://doi.org/10.1103/PhysRevLett.128.203603} {\bibfield
   {journal} {\bibinfo  {journal} {Physical Review Letters}\ }\textbf {\bibinfo
  {volume} {128}},\ \bibinfo {pages} {203603} (\bibinfo {year}
  {2022})}\BibitemShut {NoStop}%
\bibitem [{\citenamefont {Agarwal}\ \emph {et~al.}(2023)\citenamefont
  {Agarwal}, \citenamefont {Rai},\ and\ \citenamefont
  {Mondal}}]{agarwal2023quantum}%
  \BibitemOpen
  \bibfield  {author} {\bibinfo {author} {\bibfnamefont {K.}~\bibnamefont
  {Agarwal}}, \bibinfo {author} {\bibfnamefont {H.}~\bibnamefont {Rai}},\ and\
  \bibinfo {author} {\bibfnamefont {S.}~\bibnamefont {Mondal}},\ }\bibfield
  {title} {\bibinfo {title} {Quantum dots: an overview of synthesis,
  properties, and applications},\ }\href
  {https://doi.org/10.1088/2053-1591/acda17} {\bibfield  {journal} {\bibinfo
  {journal} {Materials Research Express}\ }\textbf {\bibinfo {volume} {10}},\
  \bibinfo {pages} {062001} (\bibinfo {year} {2023})}\BibitemShut {NoStop}%
\bibitem [{\citenamefont {Lan}\ and\ \citenamefont
  {Ding}(2012)}]{lan2012ordering}%
  \BibitemOpen
  \bibfield  {author} {\bibinfo {author} {\bibfnamefont {H.}~\bibnamefont
  {Lan}}\ and\ \bibinfo {author} {\bibfnamefont {Y.}~\bibnamefont {Ding}},\
  }\bibfield  {title} {\bibinfo {title} {Ordering, positioning and uniformity
  of quantum dot arrays},\ }\href
  {https://doi.org/10.1016/j.nantod.2012.02.009} {\bibfield  {journal}
  {\bibinfo  {journal} {Nano Today}\ }\textbf {\bibinfo {volume} {7}},\
  \bibinfo {pages} {94} (\bibinfo {year} {2012})}\BibitemShut {NoStop}%
\bibitem [{\citenamefont {Chen}\ \emph {et~al.}(2023)\citenamefont {Chen},
  \citenamefont {Luo}, \citenamefont {Kaplan}, \citenamefont {Bawendi},
  \citenamefont {Macfarlane},\ and\ \citenamefont {Bathe}}]{chen2023ultrafast}%
  \BibitemOpen
  \bibfield  {author} {\bibinfo {author} {\bibfnamefont {C.}~\bibnamefont
  {Chen}}, \bibinfo {author} {\bibfnamefont {X.}~\bibnamefont {Luo}}, \bibinfo
  {author} {\bibfnamefont {A.~E.~K.}\ \bibnamefont {Kaplan}}, \bibinfo {author}
  {\bibfnamefont {M.~G.}\ \bibnamefont {Bawendi}}, \bibinfo {author}
  {\bibfnamefont {R.~J.}\ \bibnamefont {Macfarlane}},\ and\ \bibinfo {author}
  {\bibfnamefont {M.}~\bibnamefont {Bathe}},\ }\bibfield  {title} {\bibinfo
  {title} {Ultrafast dense dna functionalization of quantum dots and rods for
  scalable 2d array fabrication with nanoscale precision},\ }\href
  {https://doi.org/10.1126/sciadv.adh8508} {\bibfield  {journal} {\bibinfo
  {journal} {Science Advances}\ }\textbf {\bibinfo {volume} {9}},\ \bibinfo
  {pages} {eadh8508} (\bibinfo {year} {2023})}\BibitemShut {NoStop}%
\bibitem [{\citenamefont {Tidjani}\ \emph {et~al.}(2025)\citenamefont
  {Tidjani}, \citenamefont {Denora}, \citenamefont {Chan}, \citenamefont
  {Ungerer}, \citenamefont {van Straaten}, \citenamefont {Oosterhout},
  \citenamefont {Stehouwer}, \citenamefont {Scappucci},\ and\ \citenamefont
  {Veldhorst}}]{tidjani2025threedimensional}%
  \BibitemOpen
  \bibfield  {author} {\bibinfo {author} {\bibfnamefont {H.}~\bibnamefont
  {Tidjani}}, \bibinfo {author} {\bibfnamefont {D.}~\bibnamefont {Denora}},
  \bibinfo {author} {\bibfnamefont {M.}~\bibnamefont {Chan}}, \bibinfo {author}
  {\bibfnamefont {J.~H.}\ \bibnamefont {Ungerer}}, \bibinfo {author}
  {\bibfnamefont {B.}~\bibnamefont {van Straaten}}, \bibinfo {author}
  {\bibfnamefont {S.~D.}\ \bibnamefont {Oosterhout}}, \bibinfo {author}
  {\bibfnamefont {L.}~\bibnamefont {Stehouwer}}, \bibinfo {author}
  {\bibfnamefont {G.}~\bibnamefont {Scappucci}},\ and\ \bibinfo {author}
  {\bibfnamefont {M.}~\bibnamefont {Veldhorst}},\ }\href@noop {} {\bibinfo
  {title} {A three-dimensional array of quantum dots}} (\bibinfo {year}
  {2025}),\ \bibinfo {note} {arXiv:2512.01634v1},\ \Eprint
  {https://arxiv.org/abs/2512.01634} {arXiv:2512.01634 [cond-mat.mes-hall]}
  \BibitemShut {NoStop}%
\bibitem [{\citenamefont {Katsumi}\ \emph {et~al.}(2025)\citenamefont
  {Katsumi}, \citenamefont {Takada}, \citenamefont {Jelezko},\ and\
  \citenamefont {Yatsui}}]{katsumi2025hybrid}%
  \BibitemOpen
  \bibfield  {author} {\bibinfo {author} {\bibfnamefont {R.}~\bibnamefont
  {Katsumi}}, \bibinfo {author} {\bibfnamefont {K.}~\bibnamefont {Takada}},
  \bibinfo {author} {\bibfnamefont {F.}~\bibnamefont {Jelezko}},\ and\ \bibinfo
  {author} {\bibfnamefont {T.}~\bibnamefont {Yatsui}},\ }\bibfield  {title}
  {\bibinfo {title} {Recent progress in hybrid diamond photonics for quantum
  information processing and sensing},\ }\href
  {https://doi.org/10.1038/s44172-025-00398-2} {\bibfield  {journal} {\bibinfo
  {journal} {Communications Engineering}\ }\textbf {\bibinfo {volume} {4}},\
  \bibinfo {pages} {85} (\bibinfo {year} {2025})}\BibitemShut {NoStop}%
\bibitem [{\citenamefont {Zanner}\ \emph {et~al.}(2022)\citenamefont {Zanner},
  \citenamefont {Orell}, \citenamefont {Schneider}, \citenamefont {Albert},
  \citenamefont {Oleschko}, \citenamefont {Juan}, \citenamefont {Silveri},\
  and\ \citenamefont {Kirchmair}}]{zanner2022darkstate}%
  \BibitemOpen
  \bibfield  {author} {\bibinfo {author} {\bibfnamefont {M.}~\bibnamefont
  {Zanner}}, \bibinfo {author} {\bibfnamefont {T.}~\bibnamefont {Orell}},
  \bibinfo {author} {\bibfnamefont {C.~M.~F.}\ \bibnamefont {Schneider}},
  \bibinfo {author} {\bibfnamefont {R.}~\bibnamefont {Albert}}, \bibinfo
  {author} {\bibfnamefont {S.}~\bibnamefont {Oleschko}}, \bibinfo {author}
  {\bibfnamefont {M.~L.}\ \bibnamefont {Juan}}, \bibinfo {author}
  {\bibfnamefont {M.}~\bibnamefont {Silveri}},\ and\ \bibinfo {author}
  {\bibfnamefont {G.}~\bibnamefont {Kirchmair}},\ }\bibfield  {title} {\bibinfo
  {title} {Coherent control of a multi-qubit dark state in waveguide quantum
  electrodynamics},\ }\href {https://doi.org/10.1038/s41567-022-01527-w}
  {\bibfield  {journal} {\bibinfo  {journal} {Nature Physics}\ }\textbf
  {\bibinfo {volume} {18}},\ \bibinfo {pages} {538} (\bibinfo {year}
  {2022})}\BibitemShut {NoStop}%
\end{thebibliography}
\end{document}